\documentclass[a4paper,11pt]{article}

\pdfoutput=1

\usepackage{jcappub} 
\usepackage{mathtools}
\usepackage{comment}
\usepackage{enumerate}
\usepackage{mathrsfs}
\usepackage{units}
\usepackage{graphicx,amsfonts,amssymb,amsthm,amsmath,psfrag}
\usepackage[toc,page]{appendix}
\usepackage{subfigure}
\usepackage{slashed}
\usepackage{mathtools}
\usepackage{tikz}
\usepackage{booktabs}
\usepackage{siunitx}
\usepackage{bm}         
\usepackage{color}
\usepackage{amssymb}
\usepackage{amsmath}     
\usepackage{psfrag}
\usepackage{graphicx}
\usepackage{epstopdf}
\usepackage{dcolumn}
\usepackage{bm}
\usepackage[utf8]{inputenc}
\usepackage{enumitem}
\usepackage{subfigure}
\usepackage{braket}
\usepackage[color=orange]{todonotes}
\hypersetup{linkcolor=blue,filecolor=blue,urlcolor=blue,citecolor=blue}
\usepackage[normalem]{ulem}	
\usepackage{bbding}
\usepackage{amsmath}

\newcommand{\cn}{\mathrm{cn}}

\title{
\huge{Hubble-induced phase transitions:\\ Walls are not forever }}
\author[a]{Dario Bettoni,}
\emailAdd{bettoni@usal.es}  
\author[b]{Javier Rubio}
\emailAdd{javier.rubio@helsinki.fi} 
\vspace{5cm}
\affiliation[a]{Departamento de F\'isica Fundamental and IUFFyM, Universidad de Salamanca, \\Plaza de la Merced S/N, E-37008  Salamanca, Spain}
\affiliation[b]{Department of Physics and Helsinki Institute of Physics, \\  PL 64, FI-00014 University of Helsinki, Finland} 

\abstract{The interplay between non-minimally coupled scalar fields and a kinetic-dominated era following the end of inflation triggers the spontaneous symmetry breaking of internal symmetries and the subsequent evolution of the fields towards large expectation values. We present here a detailed analysis of the associated dynamics in quintessential inflation scenarios involving a non-minimally coupled $Z_2$-symmetric spectator field. By analytically following the evolution of the spectator field fluctuations at early times, we characterize the formation of classical, homogeneous and spatially-localized field configurations separated by domain walls. The life expectancy of these dividing barriers is set by the scale of inflation, the non-minimal coupling and self-interactions of the spectator field and potentially, but not necessarily, the duration of the heating stage. For most of the parameter space, the domain walls are doomed to disappear before big bang nucleosynthesis. Potential phenomenological consequences of the scenario are discussed.}

\preprintnumber{HIP-2019-36/TH}
\begin{document}
\maketitle

\section{Introduction} \label{sec:intro}

Spontaneous symmetry breaking is an essential component of modern particle physics theories. Among the different symmetry breaking patterns that could take place in Nature, those involving discrete symmetries are usually understood as problematic in a decelerating Universe \cite{Zeldovich:1974uw}. The difficulties are related to the formation of domain walls among causally disconnected regions picking up different degenerate values when these become causally connected. As compared to standard matter or radiation components, the energy of these topological defects redshifts very slowly with the Universe expansion, meaning that, independently of their initial abundance, they will become the dominant energy contribution at sufficiently long times.  This unwanted cosmological outcome could be avoided by a suitable choice of initial conditions \cite{Coulson:1995nv} or if the discrete symmetry is anomalous \cite{Preskill:1991kd} or explicitly broken by higher-dimensional operators \cite{Rai:1992xw}, making the domain walls unstable. 

In this paper, we study the symmetry breaking dynamics of a $Z_2$-symmetric spectator field non-minimally coupled to gravity within the non-oscillatory quintessential inflation paradigm. The combination of a non-minimal coupling to gravity with the kinetic domination period ubiquitously appearing in these runaway scenarios translates into a Hubble-induced spontaneous symmetry breaking of the $Z_2$ symmetry and the subsequent displacement of the spectator field towards new time-dependent minima of the potential appearing at large field values.\footnote{We note that similar Hubble-induced mechanisms have been advocated as a solution to the domain wall problem in a supergravity context~\cite{McDonald:1997vy,Mazumdar:2015dwd}. We emphasize, however, that, beyond the origin of the Hubble-induced corrections, these scenarios display some conceptual differences with the model under consideration. First, the displacement of the fields towards large field values takes place during inflation, rather than within a post-inflationary kinetic-dominated era. Second, the domain walls in our setting are guaranteed to disappear independently of the heating and thermalization details and without the use of additional fields or symmetry breaking patterns. Indeed, it is precisely the heating stage what triggers their decay. Third, our defects, despite being short-lived, may leave some cosmological imprint in the form of a stochastic gravitational-wave background \cite{Bettoni:2018pbl}.} The transition between the old and new ground states takes place through a tachyonic or spinodal instability where the long-wavelength fluctuations of the spectator field become significantly enhanced, leading to the formation of homogeneous and localized spatial regions separated by domain wall configurations. The growth of these topological defects will stop when the total frequency of fluctuations becomes positive, either because the Universe is heated through the production of relativistic degrees of freedom in an additional sector of the theory or because the self-interactions of the spectator field itself become important. Whatever happens first, the symmetry of the spectator will be \textit{effectively} restored at that time, leading to the rapid annihilation of the domain wall configurations. 
 
This manuscript is organized as follows. After presenting an overview of the model in Section \ref{sec:overview}, we reduce it to a simple tractable form in Section \ref{sec:spectator}. The evolution of the non-minimally coupled spectator field in the absence of fluctuations is considered in Section \ref{sec:homogeneous}, leaving for Section \ref{sec:inhomogeneous} the more realistic treatment including quantum fluctuations.  Following the phase separation process in real time, we will determine the statistical properties of the emerging domains. The decay of these topological defects after symmetry restoration is considered in Section \ref{sec:decay}. Finally, our conclusions are presented in Section \ref{sec:conclusions}.

\section{The global picture}\label{sec:overview}

Quintessential inflation scenarios \cite{Peebles:1998qn,Spokoiny:1993kt} are usually formulated in terms of a canonical scalar field subject to a runaway potential, with different potential shapes giving rise to slightly different predictions.\footnote{Several formulations involving non-canonical kinetic terms have been also considered in the literature, see for instance Refs.~\cite{Wetterich:1987fm,Wetterich:1994bg,Wetterich:2014gaa,Hossain:2014xha,Rubio:2017gty,Dimopoulos:2017zvq,Dimopoulos:2017tud,Akrami:2017cir,Garcia-Garcia:2018hlc}.} In spite of the quantitative differences among these models, the absence of a potential minimum  leads, \textit{almost generically},\footnote{A potential exception are warm quintessential inflation scenarios \cite{Dimopoulos:2019gpz,Rosa:2019jci,Lima:2019yyv}, where the radiation domination era starts immediately after inflation.} to the appearance of a kinetic-dominated era soon after the end of inflation.  The existence of this unusual expansion epoch may have a strong impact on the dynamics. In particular, the onset of radiation domination does \textit{no} longer require the total depletion of the inflaton condensate since, whatever the efficiency of the heating mechanism at the end of inflation \cite{Ford:1986sy,Felder:1999pv,Rubio:2017gty}, 
the created particles will always become the dominant energy component at sufficiently long times. The interplay of kinetic domination with non-minimally coupled matter fields is also dramatic since it triggers the spontaneous symmetry breaking of internal symmetries and the subsequent evolution of the fields towards large expectation values. In what follows, we will study this transition in detail. 
We will focus here on a simple scenario involving a real spectator field $\chi$ in order to expose the essential physics in a clear as possible manner, although some of our considerations may be of more general interest.

The total action of the considered model takes the form
\begin{equation}\label{eq:S}
S=\int d^4 x \sqrt{-g} \left[\frac{M_P^2}{2}R+{\cal L}_\phi+{\cal L}_\chi\right]\,,
\end{equation}
with $M_P=2.44 \times 10^{18}$ GeV the reduced Planck mass and ${\cal L}_{\phi}$ the Lagrangian density of a quintessential inflation field $\phi$ accounting simultaneously for the early- and late-time accelerated expansion of the Universe. Although the precise form of the spectator field Lagrangian density ${\cal L}_\chi$  will not play a central role in our conclusions, we will assume it to be $Z_2$-symmetric and to involve only dimension-4 operators, namely
\begin{equation}\label{eq:lagchi}
{\cal L}_\chi=-\frac{1}{2}\partial^\mu\chi \partial_\mu \chi-\frac{1}{2}\xi R  \chi^2-\frac{\lambda}{4}\chi^4\,,
\end{equation}
where we have intentionally omitted a potential bare mass contribution that will not play any essential role in the following discussions.

We are interested in studying the above scenario in a semiclassical test-field approximation in which the backreaction of the spectator field on the spacetime dynamics can be safely neglected. Varying the action \eqref{eq:S} with respect to $\chi$ and particularizing the result for a \textit{fixed} flat Friedmann--Lema\^itre--Robertson--Walker metric $g_{\mu\nu}={\rm diag}(-1,a^2(t)\,\delta_{ij})$ with scale factor $a(t)$, we obtain the Klein--Gordon equation
\begin{equation}\label{eq:eom_chit}
\ddot{\chi}+ 3H\dot{\chi}-a^{-2}\nabla^2\chi+ \xi R \chi  +\lambda \chi^{3}= 0 \,,\,
\end{equation}
with the dots denoting derivatives with respect to the coordinate time $t$, $H=\dot a/a$ the Hubble rate and 
\begin{equation}\label{RofH}
R=3(1-3w) H^2
\end{equation}
the Ricci scalar associated to the global equation of state parameter $w$.  In order to ensure the consistency of the procedure, we will demand the contribution of the $\chi$ field in Eq.~\eqref{RofH} to stay subdominant with respect to the inflationary counterpart, such that the effective equation of state of the Universe can be well approximated by that of the inflaton condensate. Under this assumption, the mass of the $\chi$ field is positive definite during inflation ($w=-1$, $R=12 H^2$), making the spectator field heavy for $\xi\gtrsim 1/12$ and suppressing the generation of isocurvature perturbations \cite{Bettoni:2018utf,Bettoni:2018pbl}. Note, however, that as soon as the kinetic-dominated epoch starts, it induces a change of sign in the Ricci scalar ($w=1$, $R=-6H^2$), leading to the spontaneous breaking of the $Z_2$ symmetry and the subsequent evolution of the spectator field towards the new \textit{time-dependent} \textit{minima} of the potential, located now at large field values 
\begin{equation}\label{eq:chimin}
    |\chi_{\rm min}|= \left(\frac{6\xi}{\lambda}\right)^{1/2}H\,.
\end{equation}
 This Hubble-induced spontaneous symmetry breaking (SSB), initially considered in Refs.~\cite{Figueroa:2016dsc,Nakama:2018gll,Dimopoulos:2018wfg} and explored by the present authors in the context of Affleck--Dine baryogenesis \cite{Bettoni:2018utf} and gravitational wave production \cite{Bettoni:2018pbl}, was recently advocated as a heating mechanism \cite{Opferkuch:2019zbd}.

\section{Spectator field dynamics}\label{sec:spectator}

In order to study the dynamics of the spectator field during the symmetry broken phase, we will parametrize the evolution of the scale factor and the Hubble rate as 
\begin{equation}\label{eq:backkin}
a(t)= a_{\rm kin}
\left[1+3 H_{\rm kin} \left(t-t_{\rm kin})\right)\right]^{1/3}\,, \hspace{20mm}  H(t)=\frac{H_{\rm kin}}{1+3H_{\rm kin}(t-t_{\rm kin})}\,, 
\end{equation}
with $a_{\rm kin}$ and $H_{\rm kin}$ the corresponding values of these quantities at the onset of kinetic domination, arbitrarily defined at $t=t_{\rm kin}$. Additionally, we will recast the equations of motion \eqref{eq:eom_chit} in terms of a convenient set of variables
\begin{equation}
\chi\to Y=\frac{a}{a_{\rm kin}}\frac{\chi}{\,\chi_*}\,, \hspace{15mm} \vec x \to   \vec y \equiv \, a_{\rm kin} \chi_* \,\vec x \,,  \hspace{15mm} t\to z \equiv \, a_{\rm kin}\chi_* \, \tau\,,
\end{equation}
with $\chi_*\equiv \sqrt{6\xi} H_{\rm kin}$ and $\tau\equiv \int dt/a$ the conformal time. 
In terms of these quantities, the spectator field action $S_\chi=\int d^4 x \sqrt{-g}\,{\cal L}_\chi$ becomes a ``Minkowski-like" action,
\begin{equation}\label{actionchi}
S_\chi = \int d^3{\vec y} \,dz \,\left[\frac12(Y')^2 -\frac12(\nabla Y)^2 - \frac12 (6\xi-1)(\mathcal{H}^2+\mathcal{H}')Y^2-\frac{\lambda}{4}\, Y^4\right] \,,
\end{equation}
with the primes denoting derivatives with respect to the dimensionless conformal time $z$ and $\mathcal{H}(z)\equiv a'(z)/a(z)=a(z) H(z)$ the comoving Hubble rate.  Taking into account the relations 
\begin{equation}\label{az}
a(z)=a_{\rm kin}\left(1+\frac{z}{\nu}\right)^{1/2}\,, \hspace{15mm}   \mathcal{H}(z)=\frac{1}{2(z+\nu)}\,,
\end{equation}
with 
\begin{equation}
\nu\equiv \sqrt{\frac{3\xi}{2}}
\end{equation}
an integration constant ensuring that $z(t_{\rm kin})=0$,  the associated Klein--Gordon equation \eqref{eq:eom_chit} can be written as
\begin{equation}\label{EDOY}
Y''-\nabla^2\, Y-M^2(z) Y+ \lambda\, Y^3=0\,.
\end{equation}
All the dependence on the non-minimal coupling is encoded in the time-dependent mass term
\begin{equation}
M^2(z)\equiv \frac{\nu^2-1/4}{( z+\nu )^2} \,,
\end{equation}
which, as expected, vanishes identically in the conformal limit $\xi\to 1/6$ ($\nu\to 1/2$).
In the following sections, we will study the equation of motion \eqref{EDOY} in detail, providing approximate analytical solutions for the evolution of the spectator field $\chi$ during kinetic domination and the fluctuations generated by its dynamics.  

\section{The homogeneous approximation} \label{sec:homogeneous} 

\begin{figure}
    \centering
    \includegraphics[scale=.495]{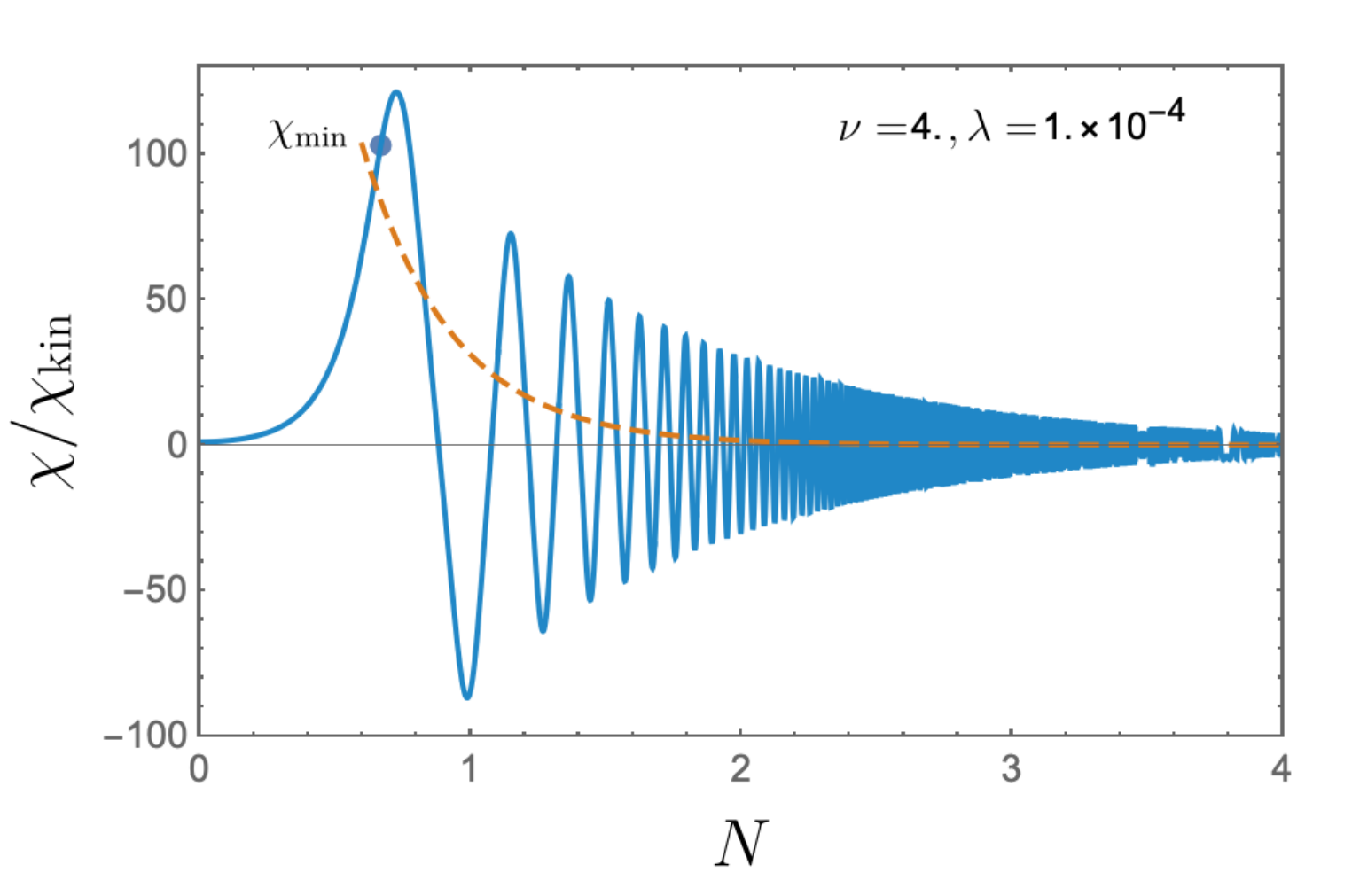}
     \includegraphics[scale=.495]{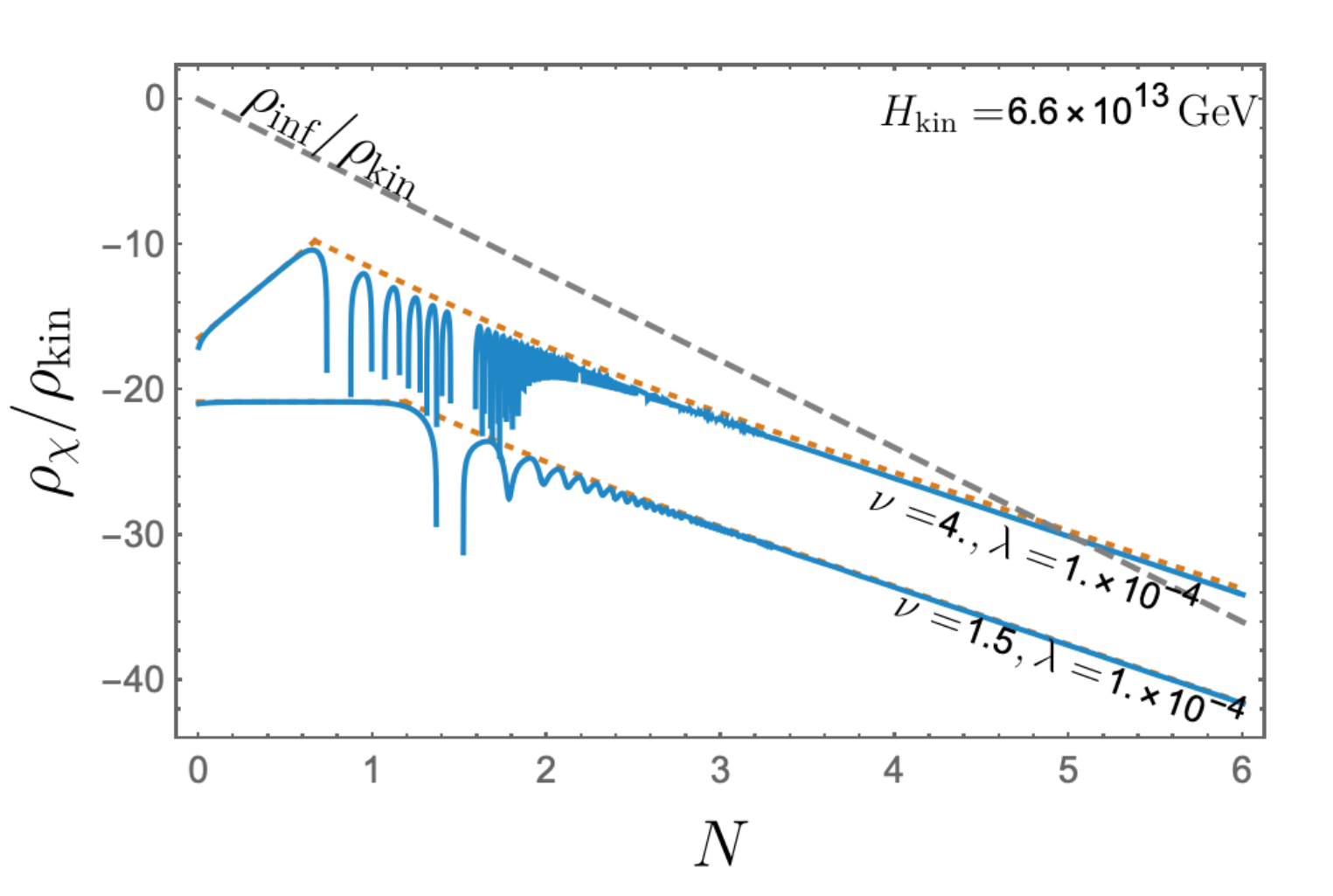}
    \caption{(Left) Evolution of the spectator field $\chi$ (solid curve) and the position of the potential minimum (dashed red curve) following from the numerical solution of Eq.~\eqref{EDOY} with $H_{\rm kin}= 10^{13}$ GeV and vanishing initial fluctuations. 
   Here,  $N\equiv \ln \left(a/a_{\rm kin}\right) =1/2 \ln \left(1+z/\nu\right)$  denotes the number of $e$-folds after the onset of kinetic domination, with the dot indicating the value at $z_b$, $N(z_b)$. (Right) Evolution of the energy density stored in the spectator field $\chi$ in the absence of quantum fluctuations (solid curve) as compared to the analytical solution \eqref{rhoapprox1} (dotted red curve) and the background energy density $\rho_\phi$ (dashed gray curve) for different values of the non-minimal coupling $\xi$. All quantities are normalized to the initial energy density. 
  Note that soon after the end of the tachyonic phase the energy density scales as radiation, similar to the result in Ref.~\cite{Opferkuch:2019zbd}.}
    \label{fig1}
\end{figure}

In order to the facilitate the comparison with previous studies in the literature \cite{Figueroa:2016dsc,Dimopoulos:2018wfg,Opferkuch:2019zbd} and highlight the limitations of those treatments, let us start discussing the dynamics of the system in the absence of quantum fluctuations. In this limit, the spectator field is described as a fully coherent state with initial conditions $\chi(t_{\rm kin})$  and $\dot \chi(t_{\rm kin})=H_{\rm kin}\,\chi(t_{\rm kin})$ or, equivalently, $Y(0)$ and $dY/d z\vert_{0} =Y(0)/\nu $.
The evolution of this condensate during the broken phase can be determined by solving the equation of motion \eqref{EDOY} in the zero-gradient approximation, 
\begin{equation}\label{EDOY1}
Y''- M^2(z)  Y+\lambda \, Y^3=0\,.
\end{equation}
As illustrated in Fig.~\ref{fig1}, the Hubble-induced tachyonic instability triggers a rapid growth of the spectator field which is eventually stopped by the cubic self-interaction term. After a transition stage where the effects of the non-minimal coupling become negligible, the field reaches an asymptotic regime where its energy density decays as radiation. A simple but sufficiently accurate description of this evolution can be obtained by matching the growing and oscillatory behaviours at the time 
\begin{equation}\label{eq:z_breaking}
\frac{z_{\rm b}}{\nu}=\left(\frac{8}{\lambda Y^2(0)}\frac{2\nu-1}{2\nu+1}\right)^{1/(2\nu+3)}-1\,,
\end{equation} 
at which the quadratic and quartic terms in the $Y$-field potential become equal.
Neglecting the decaying mode at $z\leq z_b$, we get 
\begin{equation}\label{Yapprox}
Y(z)\simeq \begin{cases}
\frac{2\nu+1}{4\nu}Y(0)\left(1+\frac{z}{\nu}\right)^{\nu+1/2} \hspace{24mm} {\rm for} \hspace{5mm} z\leq z_b\,, \\
Y_{\rm osc}\, \cn\left[\sqrt{\lambda}\,Y_{\rm osc} \,( z- z_{\rm osc}),\frac{1}{\sqrt{2}}\right]  \hspace{12mm} {\rm for} \hspace{5mm} z>z_b\,,
\end{cases}    
\end{equation}
with  $\cn$ denoting the Jacobi elliptic cosine, 
\begin{eqnarray}\label{Ymax}
Y_{\rm osc} =  \frac{Y(z_{b})}{\sqrt{2}}\left[1 + \left(1+\frac{2}{\beta^2}\frac{2\nu+1}{2\nu-1}\right)^{1/2}\right]^{1/2}
\end{eqnarray}
the amplitude of the field at the time 
\begin{eqnarray}\label{zmax}
z_{\rm osc}=z_b+\frac{1}{\,\beta\sqrt{\lambda} Y_{\rm osc}}\arccos\left(\frac{Y(z_b)}{Y_{\rm osc}}\right)
\end{eqnarray}
at which the oscillatory part reaches its maximum ($z_{\rm osc}>z_b$) and $\beta=2\pi/T_x=0.8472$ a numerical factor associated with the leading frequency of oscillation in the elliptic cosine series, 
\begin{equation}
\cn\Bigl(x,  \frac{1}{ \sqrt{2}}\Bigr)=\frac{8\pi \sqrt{2}}{T_x}\sum_{n=1}^{\infty} \frac{e^{-\pi(n-1/2)}}{1+e^{-\pi(2n-1)}}\cos \frac{2\pi (2n-1) x}{T_x} \,, 
\end{equation}
with $T_x\approx 7.416$ the oscillation period in $x$ units.

\begin{figure}
    \centering
    \includegraphics[scale=.55]{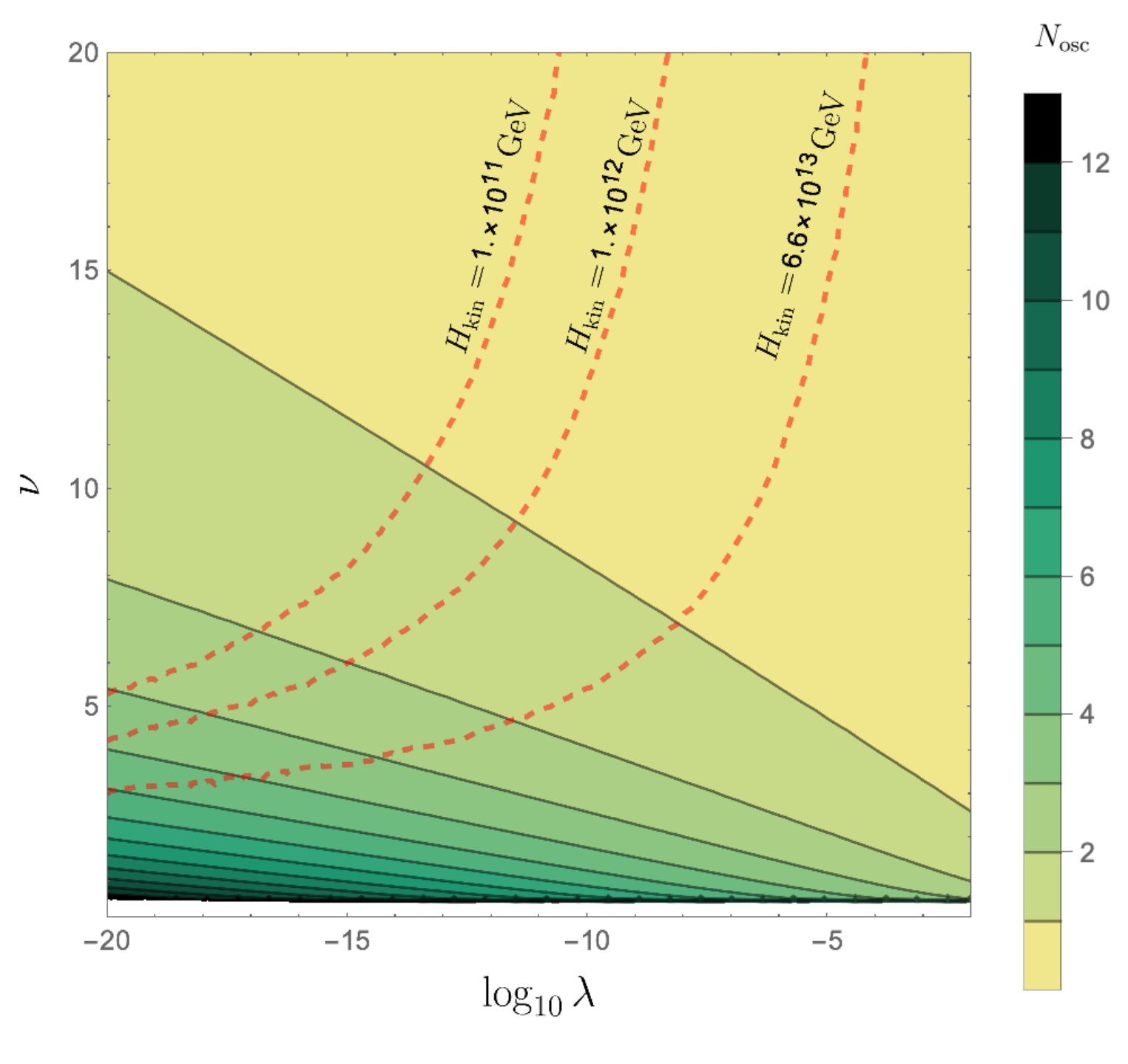}
    \caption{ Level curves for the number of $e$-folds $N_{\rm osc}=1/2\ln \left(1+z_{\rm osc}/\nu\right)$ needed for the oscillatory phase to begin as a function of the model parameters $\nu$ and $\lambda$. The dashed lines correspond to a $10\%$ spectator field correction to the Planck mass for three values of the Hubble rate at the onset of kinetic domination. Note that higher values of this quantity tend to reduce the available parameter space.}
    \label{fig:Nosc_lambda_nu}
\end{figure}

Using the analytical solution \eqref{Yapprox} we can  characterize the average evolution of the spectator field energy density 
\begin{equation}\label{rhochi}
\rho_\chi= \rho_Y {\cal X}\,, \hspace{7mm} \rho_Y\equiv \frac{1}{2}Y'^2
-\frac{1}{2}\left(6\xi-1\right)\mathcal{H}^2Y^2 + \frac{\lambda}{4}Y^4+\frac{1}{2}\left(6\xi-1\right)
    \mathcal{H}(Y^2)'
\end{equation}
during the tachyonic ($z\leq z_b$) and oscillatory phases ($z\geq z_b$), namely 
\begin{equation}\label{rhoapprox1}
\bar \rho^{\,(T)}_\chi = \,c_1 \left(\frac{a_{\rm kin}}{a}\right)^{2(1-2\nu)}   \mathcal{H}^4(0) {\cal X} \,, \hspace{15mm} \bar \rho_\chi^{\,(O)} \simeq 
\,c_2\,   \mathcal{H}^4(0) {\cal X} \,,
\end{equation}
with 
\begin{equation}
c_1\equiv (2\nu+1)^3 \nu^2  Y^2(0)\,,  \hspace{15mm}     
c_2=\frac{16\,\lambda \nu^5}{ 2\nu-1}Y^4(z_b)\,,
\end{equation}
and ${\cal X}\equiv \, \chi_*^4\left(a_{\rm kin}/a\right)^4$ a units' rescaling factor that drops when evaluating physically-relevant dimensionless ratios. 
Interestingly enough, even though the field $\chi$ is always monotonically increasing during the tachyonic phase, its energy density at $z\leq z_b$ can increase ($\nu>3/2$), decrease ($1/2<\nu<3/2$) or even stay constant if the growth of the scalar field is exactly compensated by the Universe expansion ($\nu=3/2$). On the other hand, the energy density beyond the transition value $z_b$ decreases as radiation (up to a residual ${\cal O}(a^{-6})$ correction associated with the aforementioned transient stage).  This scaling agrees with that found in Ref.~\cite{Opferkuch:2019zbd} where it was recently shown that the rapid decrease of the effective minimum depth  $\Delta = -9\xi^2/\lambda H^4 $ towards the origin of the potential precludes the trapping of the scalar field at large field values, allowing it to oscillate around the origin of the potential 
The number of $e$-folds $N_{\rm osc}=1/2\ln \left(1+z_{\rm osc}/\nu\right)$ needed for this oscillatory phase to begin can be estimated by combining the scale factor evolution in \eqref{az} with Eqs.~\eqref{eq:z_breaking}-\eqref{zmax}. As shown in Fig.~\ref{fig:Nosc_lambda_nu}, this quantity can be rather large for part of the parameter space, but, once the oscillation phase is achieved, the spectator field energy density is guaranteed to become dominant over the rapidly decreasing inflaton counterpart, $\rho_\phi \propto a^{-6}$, leading eventually to a radiation dominated epoch \cite{Opferkuch:2019zbd}.
Note that during this period, the scalar field $Y$ oscillates many times per Hubble time, cf.~Fig.~\ref{fig1}.
 
 In spite of its simplicity, the assumption of having a homogeneous, fully-coherent and smoothly-evolving scalar field adopted here and in Refs.~\cite{Figueroa:2016dsc,Dimopoulos:2018wfg,Opferkuch:2019zbd} is not physically accurate. On the one hand, the spectator field is not entirely classical and quantum fluctuations around the classical solution \eqref{Yapprox} should be definitely taken into account. On the other hand, due to the exact $Z_2$ symmetry of the original action, it is equally likely for the spectator field to roll down to any of the two minima of the effective potential, meaning that its expectation value should remain zero at all times. This trivial observation has important physical consequences. In particular, although the homogeneous approximation can be useful for estimating the temporal scales in the problem, it fails to describe the formation of topological defects \cite{Bettoni:2018pbl}, which, as we will see in what follows, it is an essential part of the problem. 

\section{Beyond homogeneity and domain wall formation}\label{sec:inhomogeneous}

Performing a complete analysis of the symmetry breaking dynamics consistently accounting for the evolution of quantum fluctuations is a rather complicated task in an interacting theory like the one under consideration. In spite of this intrinsic limitation, a lot of valuable information can be extracted from the first stages of the evolution, where the self-interactions of the spectator field  can be safely neglected. In this limit ($z<z_{\rm b}$), the action \eqref{actionchi} becomes essentially quadratic, 
\begin{equation}\label{actionchiapprox}
S_Y \simeq  \int d^3{\vec y} \,dz \,\left[\frac12(Y')^2 -\frac12(\nabla Y)^2 - \frac12 M^2(z) Y^2\right] \,,
\end{equation}
making the problem exactly solvable. To perform the quantization of the real perturbation $Y(\vec y,z)$ we will follow the usual canonical quantization scheme, although path integral techniques could be of course alternatively applied. The Hamiltonian in momentum space,
\begin{equation}
H = \frac{1}{2}  \int d^3\kappa\,\left[\Pi_{\vec \kappa}(z)\,\Pi^\dagger_{\vec \kappa}(z) +
 \omega_{\kappa}^2(z)\,Y_{\vec \kappa}(z)\,Y^\dagger_{\vec \kappa}(z)\right]\,,
\end{equation}
becomes the sum of a continuous set of uncoupled harmonic oscillators with wavenumber
\begin{equation}
 \kappa\equiv  \vert \vec \kappa \vert\equiv \frac{\vert \vec k\vert }{a_{\rm kin}\chi_*} 
\end{equation}
and position and momentum operators satisfying the standard equal-time commutation relations $
\left[Y_{\vec\kappa}(z),\Pi_{\vec \kappa}(z)\right]=i \delta^3(\vec \kappa + \vec \kappa')$.
A simple inspection of the time-dependent frequency of these operators, 
\begin{equation}\label{freq}
 \omega_{\kappa}^2(z)\equiv \kappa^2-M(z)^2\,,
	\end{equation} 
reveals that the tachyonic or spinodal instability appearing at the the onset of kinetic domination affects \textit{not only} the zero mode of the spectator field but also all \textit{subhorizon} modes with momenta smaller than the Hubble-induced negative mass term $-M^2(z)$, i.e. those within a  window $\kappa_{\rm min} (z)\lesssim \kappa \lesssim \kappa_{\rm max}(z)$ with
\begin{equation}\label{eq:kapparange}
\kappa_{\rm min}(z)=\mathcal{H}(z)\,,\hspace{20mm} \kappa_{\rm max}=(4\nu^2-1)^{1/2}\kappa_{\rm min}\,.
\end{equation} 
\begin{figure}
    \centering
\includegraphics[scale=.48]{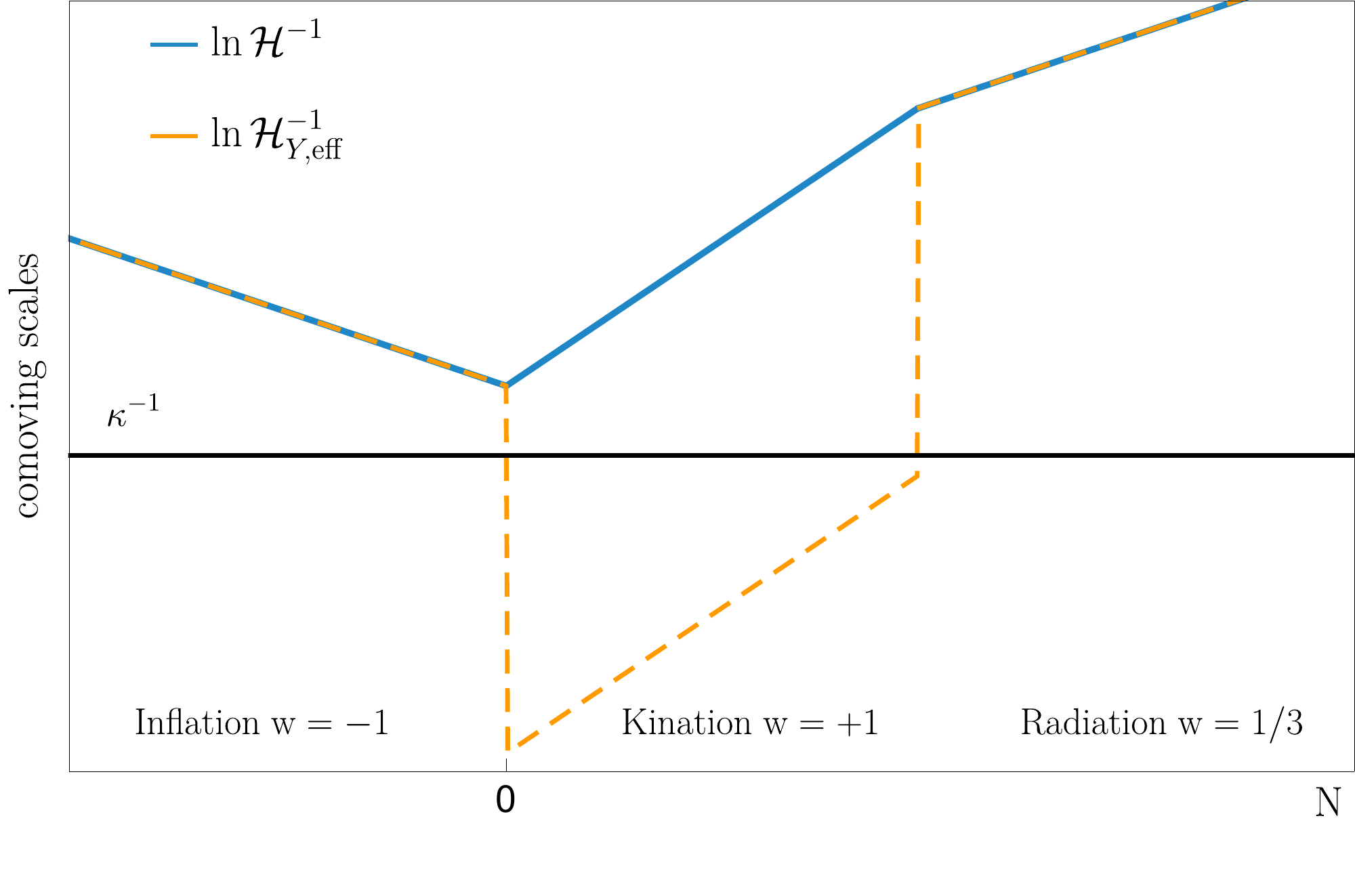}
    \caption{Schematic behaviour of ``effective horizon radius" of the Y field ${\cal H}^{-1}_{Y,{\rm eff}}$ (coinciding  during kination with the inverse mass $M^{-1}$) and the proper horizon radius ${\cal H}^{-1}$ during the different cosmological epochs. The tachyonic instability affects all scales $\kappa^{-1}$ entering the parallelogram in the center of the figure. Note that the separation between ${\cal H}^{-1}_{Y,{\rm eff}}$ and  ${\cal H}^{-1}$ during kinetic domination grows with $\nu$, making the integrated contribution of the superhorizon scales reentering the horizon ${\cal H}^{-1}$ less and less important as compared with that of modes never amplified by the inflationary dynamics.
}    \label{fig:sketch}
\end{figure}
Although this instability is reminiscent from that appearing in chaotic inflation \cite{Mukhanov:1990me}, new inflation \cite{Guth:1985ya} or tachyonic preheating scenarios \cite{Felder:2000hj,Felder:2001kt,Copeland:2001qw,Asaka:2001ez,GarciaBellido:2001cb,Copeland:2002ku,GarciaBellido:2002aj}, there are important differences with these settings: 
\begin{enumerate}
    \item  The $z$ variable in our case runs from 0 to $\infty$, as opposed to what happens in a de Sitter stage, where it runs from $-\infty$ to $0$.  This precludes the use of large argument asymptotic expansions to fix the initial conditions for the problem solutions. 
    \item   Rather than from its own background field configuration, the spectator field takes energy directly from the cosmological expansion induced by the runaway inflaton condensate.
    \item The tachyonic instability is limited in time  since the modes will eventually leave the unstable band either because of the evanescence of the tachyonic mass $-M^2(z)$ or because the full effective frequency (i.e. the one including also the omitted $\lambda Y^3$ contribution) becomes positive, cf. Fig.~\ref{fig:sketch}. 
\end{enumerate}
The physical content of the theory is encoded, as usual, in the expectation value of field operator products at equal or different spacetime points. For a zero-mean Gaussian field like the one under consideration, the only quantities needed to describe the system are the two-point expectation values 
\begin{equation}\label{exp}
\langle  v^I_{\vec\kappa}(z)\, v^J_{\vec \kappa'}(z') \rangle = \Sigma^{IJ}_\kappa(z,z')\, \delta^3({\vec \kappa}+{\vec \kappa'}) \,,
\end{equation}
with $v_{\vec \kappa}\equiv \left(\Pi_{\vec \kappa}(z),Y_{\vec\kappa}(z)\right)^T$ and the bra-kets referring to a quantum average over the initial state (Heisenberg picture). The correlation matrix $\Sigma^{IJ}_\kappa(z,z')$ in this expression can be expressed in terms of its value at any reference time $z_r$ by taking into account the Heisenberg equations, identical in form to the classical field equations
\begin{equation}
\frac{d}{dz} v_{\vec k}(z) \equiv\frac{d}{dz}  \begin{pmatrix} \Pi_{\vec \kappa}(z) \vspace{2mm} \cr Y_{\vec\kappa}(z) \end{pmatrix}= 
\begin{pmatrix}
0 & -\omega^2_{\kappa}(z)\vspace{2mm}\cr 1 &0
\end{pmatrix} 
\begin{pmatrix}\Pi_{\vec\kappa}(z) \vspace{2mm} \cr Y_{\vec \kappa}(z)
\end{pmatrix} \,.
\end{equation}
Denoting by ${\mathbf M}_{\kappa}(z)$ the evolution matrix relating the operators at times $z$ and $z_r$ (i.e. $v_{\vec \kappa}(z) = {\mathbf M}_{\kappa}(z) v_{\vec \kappa}(z_r)$), we get  \cite{Polarski:1995jg,Lesgourgues:1996jc,Kiefer:1998jk,GarciaBellido:2002aj} 
\begin{equation}\label{eq:correlationmatrix}
\Sigma_{\kappa}(z,z')={\mathbf M}_{\kappa}(z)\,\Sigma_{\kappa}(z_r,z_r)\, {\mathbf M}^T_{\kappa}(z')\,, \hspace{5mm} {\mathbf M}_{\kappa}(z)  =
\begin{pmatrix} \sqrt{\frac{2}{\kappa}}\, \textrm{Re}\, g_{\kappa}(z)&\sqrt{2\kappa}\, \textrm{Im}\, g_{\kappa}(z)\vspace{2mm}\cr  
-\sqrt{\frac{2}{\kappa}}\,\textrm{Im}\,f_{\kappa}(z) & \sqrt{2\kappa}\,\textrm{Re}\, f_{\kappa}(z) 
\end{pmatrix} \,,
\end{equation}
with $g_\kappa\equiv i f'_{\kappa}$ and $f_\kappa$ a solution of the Schr\"odinger-like differential equation
\begin{equation}\label{eq:feom}
f_\kappa '' + \omega^2_\kappa(z)\,f_\kappa = 0\,, 
\end{equation}
with initial conditions $ f_\kappa(z_r)$ and $f'_{\kappa}(z_r)$. 

The precise form of the initial correlation matrix $\Sigma_{\kappa}(z_r,z_r)$ depends on the state of fluctuations at the onset of kinetic domination and therefore on the details of the inflation-to-kination transition.  Note also that, contrary to what is usually assumed in standard heating scenarios, there is generically not a clear separation of scales in the problem at hand. Indeed, the mass of the spectator field is not significantly larger than the Hubble rate for small $\nu$ values, meaning that horizon entry effects cannot be completely ignored. To surpass these limitations we will i) consider the inflation-to-kination transition to be instantaneous as compared to the temporal evolution of the spectator field, ii) focus on a large $\nu$ limit where the impact of horizon-entry effects is minimized by increasing the separation between the horizon scale and the dominant momentum scale of the spectator field  fluctuations (cf. Fig.~\ref{fig:sketch}) and iii) assume the initial state of the system to be described by a vacuum Hermitian correlation matrix
\begin{equation}
\Sigma_{\kappa}(0,0)= \begin{pmatrix} \frac{\kappa}{2} & - \frac{i}{2}\vspace{2mm}\cr
\frac{i}{2} & \frac{1}{2\kappa}\end{pmatrix}
\end{equation}
at $z_r=0$.
Under these assumptions, the solution of the equation of motion \eqref{eq:feom} with initial conditions $ f_\kappa(0) = 1/\sqrt{2\kappa}$ and $ f'_{\kappa}(0)=-i\sqrt{\kappa/2}$ 
takes the form 
\begin{equation}\label{eq:soltach}
    f_{\kappa}(z)=\sqrt{z+\nu}\left[A_{\kappa}\, \mathcal{J}_\nu(\kappa(z+ \nu))- B_{\kappa}\,  \mathcal{Y}_\nu(\kappa(z+\nu)) \right]\,,
\end{equation}
with  $\mathcal{J_\nu}$ and $\mathcal{Y_\nu}$ the Bessel's functions of the first and second kind, 
\begin{equation}
A_{\kappa}= {\cal Y}_{\nu }(\kappa \nu) \,\delta \,, \hspace{15mm}
  B_{\kappa}= \,{\cal J}_{\nu }(\kappa
   \nu)\,\delta-\frac{f_{\kappa}(0)}{\sqrt{\nu} {\cal Y}_{\nu }(\kappa \nu)} \,, 
\end{equation}
and 
\begin{equation}
  \delta = \frac{\pi  f_{\kappa }(0)}{4\sqrt{\nu}}  \left[1-2 \nu  +2
  \nu\left(\kappa  \frac{ {\cal Y}_{\nu -1}(\kappa  \nu)}{{\cal Y}_{\nu }(\kappa \nu) }-\frac{f'_{ \kappa }(0)}{f_{\kappa }(0)}\right)\right]\,.
\end{equation}
Using this analytical solution we can evaluate the equal-time correlation matrix following from  \eqref{eq:correlationmatrix}, namely 
\begin{equation}\label{Sigmaz}
\Sigma_{\kappa}(z,z)=
\begin{pmatrix} 
|g_{\kappa}(z)|^2 &  F_\kappa(z) - \frac{i}{2}\vspace{3mm}\cr
 F_\kappa(z) + \frac{i}{2} &  |f_k(z)|^2
\end{pmatrix}\,,
\end{equation}
with 
\begin{equation}
F_\kappa(z)\equiv 
\frac12\,\langle  \Pi_{\vec \kappa}(z)\, Y_{\vec \kappa}^\dagger(z) +  Y_{\vec \kappa}(z)\, \Pi_{\vec \kappa}^\dagger(z)\rangle
= -\,\frac{i}{2}\,(g_\kappa\,f_\kappa^* - f_\kappa\,g_\kappa^*) =  {\rm Im}\, (f_\kappa^* g_\kappa)  
\end{equation}
the so-called WKB phase. 
\begin{figure}
    \centering
\includegraphics[scale=.495]{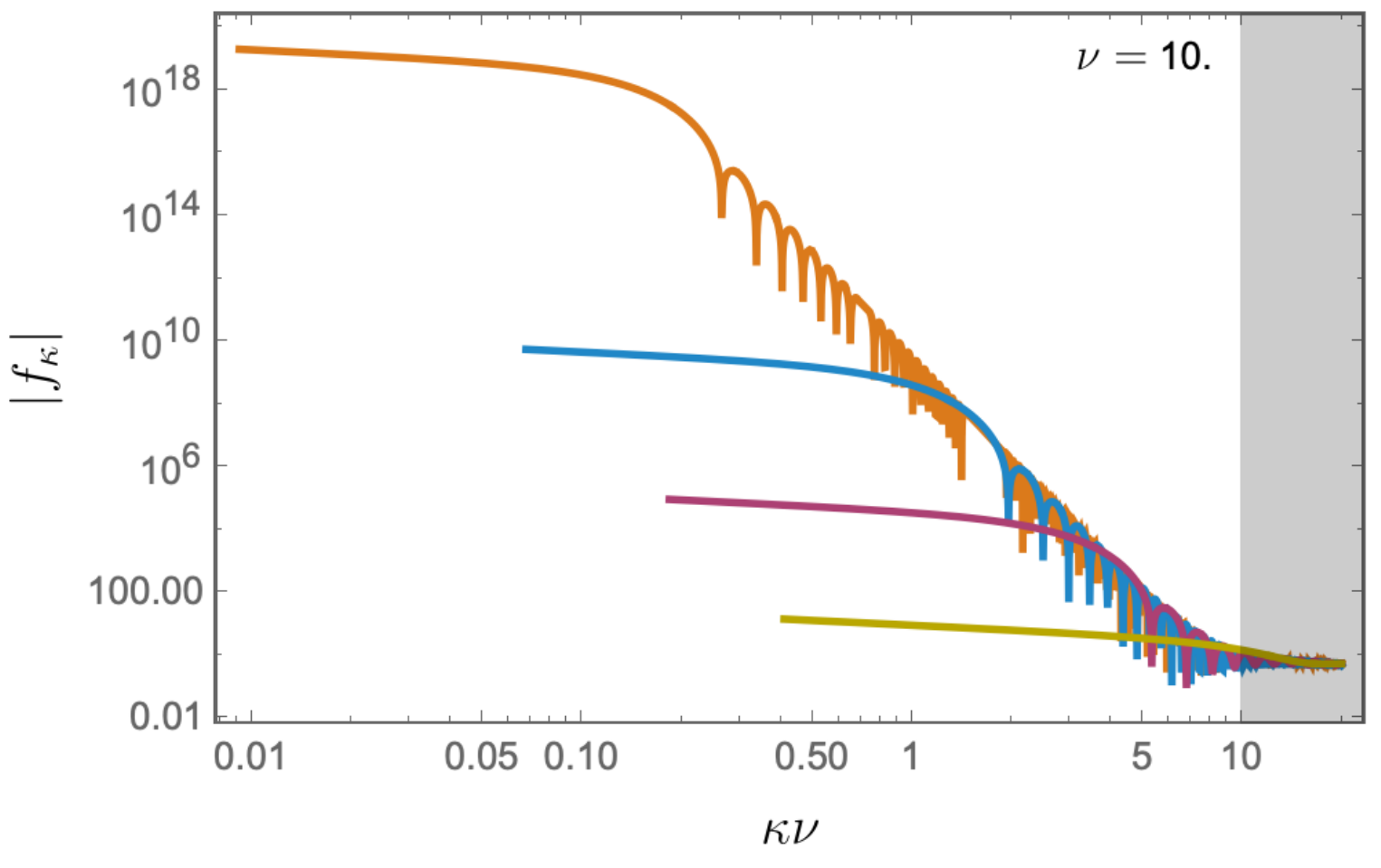}
\includegraphics[scale=.48]{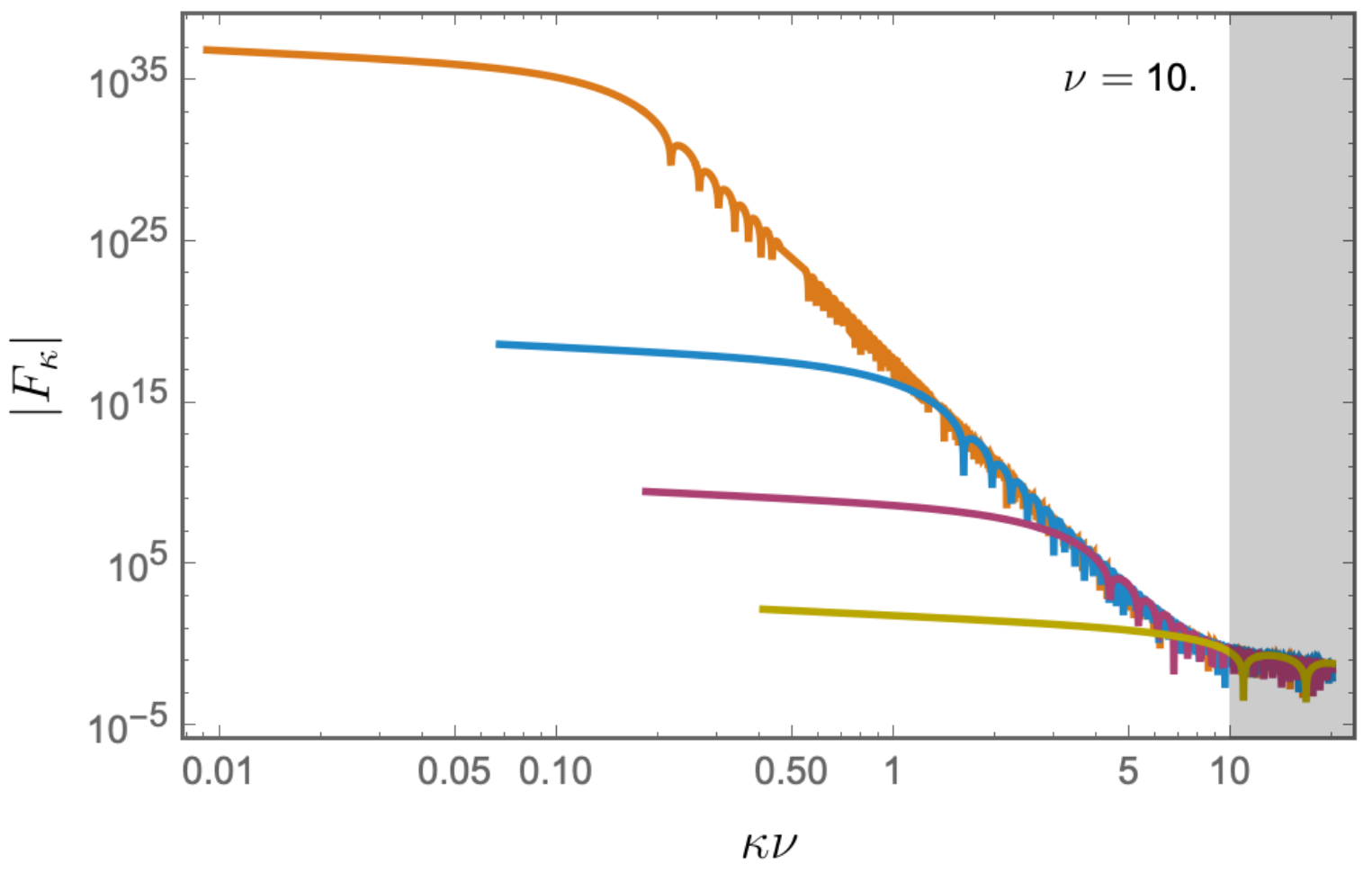}
    \caption{The functions $\vert f_\kappa\vert$ and $\vert F_\kappa\vert$ entering in the equal-time correlation matrix \eqref{Sigmaz} and following from Eq.~\eqref{eq:soltach} at $N=0.1, 0.5, 1$ and  $2$ $e$-folds (from bottom-up). The shaded region corresponds to momenta $\kappa\geq \kappa_{\rm max}(0)$. The position of the knee in the spectrum matches the value of $\kappa_{\rm max}(z)$.}    \label{fig:Ykz}
\end{figure}
The form of the different entries in this correlation matrix is illustrated in Fig.~\ref{fig:Ykz}. As clearly appreciated there, the behaviour of the functions at a given instant $z$ can be roughly divided into 3 distinct regions. First, there exists a highly-infrared amplification band at $\kappa_{\rm min}(z)<\kappa< \kappa_{\rm max}(z)$ where all modes are almost democratically amplified. Second, we can identify a strongly oscillating region $\kappa_{\rm max}(z)<\kappa<\kappa_{\rm max}(0)$ associated with modes that, although initially amplified by the tachyonic instability, are currently unaffected by it. Finally, there is a stable region  $\kappa>\kappa_{\rm max}(0)$ where modes never gets amplified.  To get some analytical understanding on this numerical result, let us note that among the two terms in Eq.~\eqref{eq:soltach}, the first one gives the largest contribution, allowing us to approximate the mode function as 
\begin{equation}\label{eq:soltachapprox}
 f_{\kappa}(z)\simeq A_{ \kappa}\, \sqrt{z+\nu} \mathcal{J}_\nu(\kappa(z+ \nu))\,.
\end{equation}
A simple description of this growing mode for the relevant \textit{subhorizon} momenta we are interesting in can be obtained by expressing it in terms of the generalized hypergeometric series $_0 F_1$ \cite{abramowitz+stegun}
\begin{equation}\label{eq:soltachapprox1}
 f_{\kappa}(z)=\frac{A_{ \kappa}\, \sqrt{z+\nu}}{\Gamma (\nu+1)} \left(\frac{\kappa  (z+\nu)}{2}\right)^{\nu} {}_0 F_1\left(\nu+1,-\left(\frac{\kappa  (z+\nu)}{2}\right)^2\right)\,.
\end{equation}
Further approximating\footnote{Note that this approximation becomes \textit{exact} in the large $\nu$ limit,
\begin{equation}
\lim_{\nu\to \infty}  \Gamma(\nu+1) \left(\frac{2}{\kappa(z+\nu)}\right)^\nu \mathcal{J}_\nu(\kappa(z+\nu)) =\exp\left(-\frac{(\kappa(z+\nu))^2}{4\nu}\right)\,.  
\end{equation}}
\begin{equation}
  \vert A_{\kappa}\vert \simeq \frac{\vert 2 \nu-1 \vert}{8}\left(\frac{2}{\kappa  \nu
   }\right)^{\nu+1/2}\Gamma(\nu)\,, \hspace{7mm}
 {}_0 F_1\left(\nu+1,-\left(\frac{\kappa  (z+\nu)}{2}\right)^2\right) 
\approx  \exp\left(-\frac14\frac{\kappa^2}{\kappa_*^2(z)}  \right)\,,
\end{equation}
with $\Gamma(\nu)$ the Euler gamma function and
\begin{equation}\label{kappatyp}
   \kappa_*(z)\equiv 2\sqrt{\nu+1}\, \kappa_{\rm min}(z)
\end{equation}
a typical momentum scale, we get 
\begin{eqnarray}\label{f2approx}
\vert f_\kappa(z) \vert^2 &\approx& \left(\frac{2 \nu-1}{4\sqrt{2}\nu}\right)^2   \left(1+\frac{z}{\nu}\right)^{2\nu+1}\frac{1}{\kappa}\exp\left(-\frac{1}{2}\frac{\kappa ^2}{\kappa^2_*(z)}\right)  \,, \\
\vert g_\kappa(z)\vert^2 &\approx& \frac{|f_\kappa(z)|^2}{2}\left(\frac{2 \nu+1}{z+\nu}-\frac{\kappa^2(z+\nu)}{1+\nu}\right)^2\,,  \\
\vert F_\kappa \vert &\approx&   \frac{\vert f_{\kappa}(z)\vert^2}{2}\left\vert\frac{2 \nu+1}{z+\nu}-\kappa^2\frac{z+\nu}{\nu+1}\right\vert\,. \label{Fapprox}
\end{eqnarray}
Note that the growth factor of fluctuations for infrared modes $\kappa\ll \kappa_*(z)$ coincides with that of the zero mode in Eq.~\eqref{Yapprox}. This rapid enhancement is intimately related to the emergence of classicality in the corresponding range of momenta. In particular, the Heisenberg uncertainty principle in Fourier space can be written as \cite{Polarski:1995jg,Kiefer:1998jk,Lesgourgues:1996jc,GarciaBellido:2002aj}
\begin{equation}
\Delta Y_\kappa^2\,\Delta \Pi_\kappa^2 = \vert F_\kappa(z)\vert^2 + \frac14 \geq {\frac14}\, \Big\vert \langle [Y_\kappa(z),\ \Pi^\dagger_\kappa(z)]\rangle\Big\vert^2\,,
\end{equation}
meaning that for $\vert F_\kappa(z)\vert \gg 1$ (or equivalently for $\langle\{Y_\kappa(z),\ \Pi_\kappa^\dagger(z)\}\rangle \gg \langle|[ Y_\kappa(z), \Pi_\kappa^\dagger(z)]|\rangle =  \hbar$), 
the ambiguity in the ordering of operators becomes quantitatively negligible. Thanks to this quantum-to-classical transition, the statistical properties of the system can be completely described in terms of a (dimensionless) power spectrum 
\begin{equation}\label{PSdef}
P_\kappa(z) \equiv  \kappa^3|f_\kappa(z)|^2
   \end{equation}
 consistent with the conservation of probabilities along classical trajectories.
 Several interesting quantities characterising this classical Gaussian random field in position space can be computed by Fourier transforming expressions of the form $\kappa^p(\kappa) P(\kappa)$, with $p$ a positive even integer.
\begin{figure}
    \centering
\includegraphics[scale=.65]{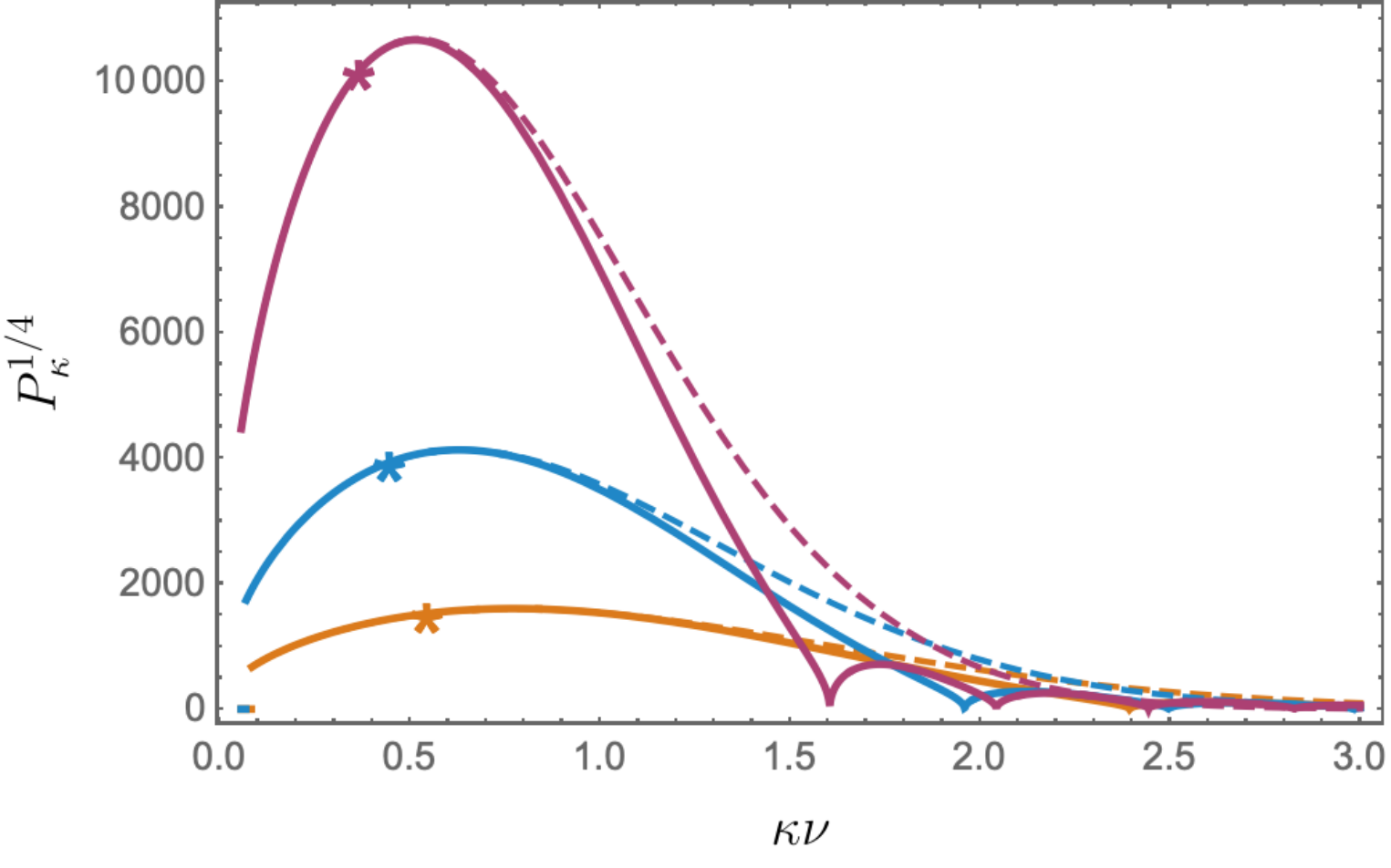}
    \caption{Comparison between the late-time behaviour of the power spectrum following from the exact solution \eqref{eq:soltach} (solid line) and the Gaussian approximation \eqref{PSapp} (dashed line) for $\nu=10$ and $N=0.9,1$ and $1.1$ $e$-folds. The markers refers to the position of $\kappa_* \nu$ at the corresponding number of $e$-folds.\label{fig:Px}}  
\end{figure}
These mode integrals are ultraviolet divergent ($\vert f_\kappa\vert^2\sim 1/\kappa$, $P(\kappa)\sim \kappa^2$ at large $\kappa$) and must be regularized in order to perform any practical computation.
A measurable quantity in the effective field theory sense can be defined, for instance, by integrating out momenta above a given cutoff scale. When doing that, quadratic and logarithmic divergences can be absorbed in the renormalized constants of the theory \cite{Boyanovsky:1993xf,Cooper:1994hr,Cooper:1996ii,Baacke:1996se,Baacke:1997rs,Boyanovsky:1998yp,GarciaBellido:2002aj}. A natural way of introducing this cutoff in our setting is to replace the mode function $\vert f_\kappa (z)\vert^2$ by its approximate form \eqref{f2approx}. This effective truncation smears out the problematic high-momentum modes leading behind an approximately Gaussian spectrum 
\begin{equation}\label{PSapp}
P_\kappa(z)  \approx \left(\frac{2 \nu-1}{4\sqrt{2}\nu}\right)^2  \left(1+\frac{z}{\nu}\right)^{2\nu+1}\kappa^2  \exp\left(-\frac{1}{2}\frac{\kappa ^2}{\kappa^2_*(z)}\right) \,,
   \end{equation}
which can be analytically integrated over. The comparison between this simple expression and that obtained from the exact solution \eqref{eq:soltach} is depicted in Fig.~\ref{fig:Px}. As clearly appreciated in this plot, soon after the beginning of the phase transition, the spectator field power spectrum develops a peak around the typical momentum scale \eqref{kappatyp} which redshifts and grows with time until eventually dominating the infrared part of the spectrum. Note that the Gaussian approximation \eqref{PSapp} accurately captures this leading behaviour, failing only in the numerically subdominant oscillatory part at large $\kappa$ values, as otherwise expected. 
 
\begin{figure}
    \centering
\includegraphics[scale=.65]{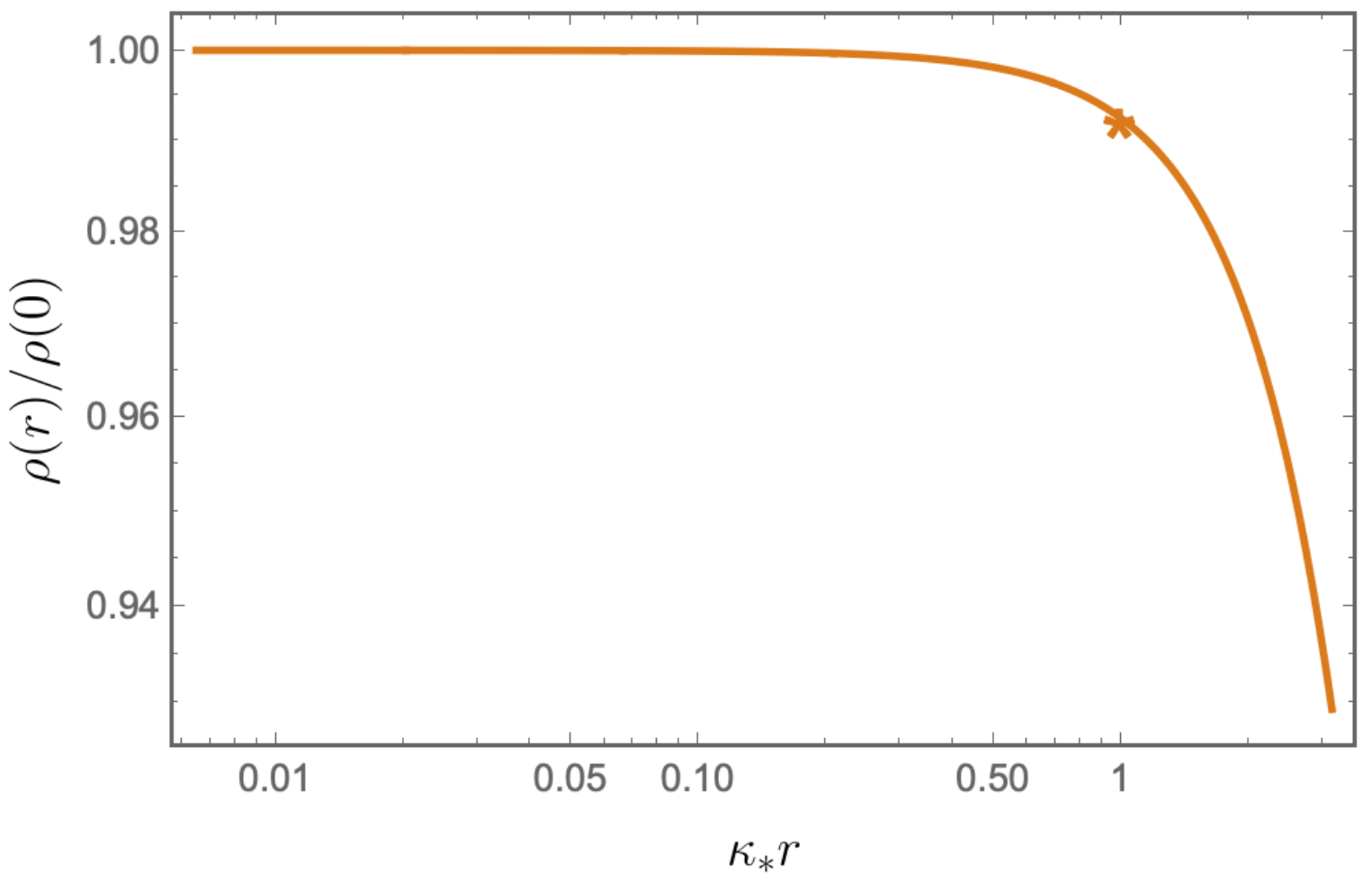}
    \caption{Density profile \eqref{eq:profile} normalized to the central density. 
    The star signals $\kappa_* r =1$.
    \label{fig:profile}}  
\end{figure}

The low-momentum structure of the power spectrum translates into the formation of localized domains in position space. The size of the correlated regions is determined by the spatial correlation function, defined as
\begin{equation}\label{eq:size0}
\zeta(\vec y,z) \equiv \left\langle Y(\vec y,z)Y(0,z)\right \rangle = \int \frac{d^3\kappa }{(2\pi)^3} \, e^{i\vec \kappa \cdot \vec y}\, \vert f_\kappa(z)\vert^2=\frac{1}{2\pi^2} \int  e^{i\vec \kappa \cdot \vec y}\, P_\kappa(z) \, d\ln \kappa\,.
\end{equation}
Taking into account Eq.~\eqref{PSapp} and reducing the 3-dimensional integral in this expression to a 1-dimensional integral by switching to spherical coordinates,  we get
\begin{equation}\label{eq:size}
\zeta(r,z) = \frac{1}{2\pi^2}
\int_0^\infty P_\kappa(z)\,j_0(\kappa r) \,d\ln \kappa \simeq \zeta(0,z)  \,G_1(\kappa_* r) \,,
\end{equation}
with $j_0(\kappa r)=\sin\,(\kappa r)/(\kappa r)$ the spherical Bessel function.\footnote{Note that we have omitted an IR cutoff at $\kappa=\kappa_{\rm min}(z)$. The error we are making is, however, small, namely of the order of percent for $\nu=10$. In fact, $\delta\zeta/\zeta=1-\exp[-1/(8(\nu+1))]$ in the $r\rightarrow 0$ limit.} Here 
\begin{equation}\label{Yrms}
 \zeta(0,z)\equiv Y^2_{\rm rms}(z)= \left(\frac{2 \nu-1}{8\pi \nu}\right)^2    \left(1+\frac{z}{\nu}\right)^{2 \nu +1} \kappa_*^2(z)
\end{equation}
stands for the square of the time-dependent dispersion determining the root mean-square perturbation (rms) and 
\begin{equation}
G_1(\kappa_* r)\equiv \frac{\sqrt{2}}{\kappa_* r} D\left(\frac{r\, \kappa_*}{\sqrt{2}}\right)  = \sqrt{\frac{\pi}{2}}\frac{1}{\kappa_* r} \exp\left(-\frac{1}{2}\kappa^2_* r^2\right) \text{erfi}\left(\frac{ \kappa_* r}{\sqrt{2}}\right)
\end{equation}
is a shape function approaching unity at $r\to 0$, with $D$ the Dawson function and  $\text{erfi}$ the imaginary error function \cite{abramowitz+stegun}. Note that, although the mean field value is zero when averaged over the entire ensemble, typical field configurations within the ensemble are non vanishing, but rather sample positive and negative values. At a given time $z$, the correlation length is dominated by wavelengths of the order of the inverse momentum $\kappa^{-1}_*(z)$ and therefore much smaller than the horizon radius for large $\nu$ values. This is also reflected in the profile of the domains, which, following similar steps to those above, can be written as \cite{Bardeen:1985tr}
\begin{equation} 
\rho(r,z) = \frac{1}{\sqrt2\,\pi} \int_0^\infty P_\kappa^{1/2}(z)\,
j_0(\kappa r)\,  d\ln\kappa\simeq \rho(0,z) G_2(\kappa_* r)\,,
\end{equation}
with 
\begin{equation}\label{eq:profile}
\rho(0,z)\equiv \pi^{1/2} \frac{ \vert 2 \nu-1\vert }{8\pi \nu}\left(1+\frac{z}{\nu}\right)^{\nu +1/2} \kappa_*(z)\,, \hspace{10mm} G_2(\kappa_* r)\equiv \frac{\sqrt{\pi}}{2}\frac{\text{erf}(\kappa_* r)}{\kappa_*r}\,,
\end{equation}   
and $\text{erf}$ the error function, cf.~Fig.~\ref{fig:profile}.

Although classical and sufficiently localized, the identification of the created topological defects in a Gaussian random field is a generically a complicated task.  The number density of domain walls can be estimated by counting the number of zeros of the classical field, given by \cite{Halperin,Liu:1992zz}
\begin{equation}\label{nzeros}
    n_{\rm walls}(z) = \frac{1}{\pi^2}\left(-\frac{\zeta''(0,z)}{\zeta(0,z)}\right) ^{3/2}\,,
\end{equation}
with 
\begin{equation}
    -\frac{\zeta''(0,z)}{\zeta(0,z)} = \frac{\int d\ln \kappa\, \kappa^2 P_\kappa(z)}{\int d\ln \kappa\, P_\kappa(z)}\,.
\end{equation}
We get 
\begin{equation}\label{eq:npeak}
n_{\rm walls}(z) = \left(\frac{2}{3}\right)^{3/2}  \frac{\kappa_*^3(z)}{\pi^2}\,.
\end{equation}

\section{Symmetry restoration and domain wall annihilation}\label{sec:decay}

The applicability of the previous section techniques is restricted to the early times at which the self-interactions of the created particles can be safely neglected ($\lambda\to 0$). A useful formalism able to account for potential backreaction effects in a time-dependent setting is the two-particle-irreducible close-time path or in-in formalism,  whose truncation at the two-loop order becomes equivalent to the time-dependent Hartree--Fock approximation. In this limit, the quartic non--linearity is replaced by $3/2 \, \lambda \langle Y^2\rangle  Y^2$, leading to the following effective frequency for the spectator field fluctuations 
\begin{equation}\label{freqback}
 \omega_{\kappa}^2(z)\equiv \kappa^2-M(z)^2+3\lambda Y^2_{\rm rms}(z) \,.
	\end{equation} 
The tachyonic instability at small momenta will stop when this modified frequency becomes positive, leading in practice to symmetry restoration. 
This symmetry restoration can be ``\textit{effective}" if the field is stabilized by its own interactions and starts oscillating with a sizable amplitude around the origin of the potential (cf.~Section \ref{sec:homogeneous}) or ``\textit{exact}" if the Universe is heated through some additional sector before the onset of non-linearities.
 
As argued in Ref.~\cite{Rubio:2017gty}, the heating effects can be parametrized by a \textit{heating efficiency} parameter 
\begin{equation}\label{thetadef}
\Theta\equiv \frac{\rho_R}{\rho_\phi}\Big\vert_{z_{\rm kin}}=\left(\frac{a_{\rm kin}}{a_{\rm rad}}\right)^2\,,
\end{equation}
with $\rho_R$ the energy density of the heating products and $a_{\rm kin}$ and $a_{\rm rad}$ the values of the scale factor at the onset of kinetic and radiation domination.  The minimal value of the heating efficiency
\begin{eqnarray}\label{GWbound2}
&&\Theta\gtrsim          10^{-16}\left(\frac{H_{\rm kin}}{10^{11}\,{\rm GeV}}\right)^2\,,
\end{eqnarray}
is restricted by the (integrated) nucleosynthesis constraint on the gravitational waves density fraction \cite{Rubio:2017gty}.
 Taking into account the scaling of the radiation and inflaton energy densities during kinetic domination, 
\begin{equation}\label{XAB}
\frac{\rho_R(a)}{\rho_\phi(a)}=\Theta\left(\frac{a}{a_{\rm kin}}\right)^2\,,
\end{equation}
we can easily determine the number of $e$-folds after the end of inflation at which the $Z_2$ symmetry is ``\textit{exactly}" restored  by heating  effects  ($a=a_{\rm rad}$, $\rho_{\rm R}(a_{\rm rad})=\rho_\phi (a_{\rm rad})$, $R\simeq 0$), 
\begin{equation}
N_{\rm exact}=  \ln \left(\frac{a_{\rm kin}}{a_{\rm rad}}\right)=\frac{1}{2}\ln \Theta\,.
\end{equation}
 This temporal scale should be compared with the ``\textit{effective}" restoration time at which the last two terms in Eq.~\eqref{EDOY} become approximately equal [$\lambda Y_{\rm rms}^2/2 \simeq M^2$], namely 
\begin{equation}\label{eq:backreacttime}
N_{\rm eff}= \frac{1}{2}  \ln  \left(1+\frac{z_{\rm eff}}{\nu}\right)  \simeq {\frac{1}{2 \nu +1}} \ln \left[\frac{8\pi\nu^2}{(2\nu-1)\sqrt{3\lambda(\nu+1)}}\left(1-\frac{1}{4\nu^2}\right)^{1/2}\right]\,.
\end{equation} 
Beyond the smallest of these quantities,
the approach presented in Section \ref{sec:inhomogeneous} becomes inaccurate and the theory cannot be longer described as a free theory where the gaussianity of the initial state is preserved by the evolution.

Immediately after symmetry restoration the Universe is divided into regions with a rather homogeneous field distribution on scales $L\lesssim \kappa^{-1}_*$, cf. Fig.~\ref{fig:profile}. Any homogeneous or perturbative approximation ignoring this fact will be unable to provide a realistic description of the problem at hand, disregarding in the process interesting physical effects such as the potential generation of gravitational waves by the walls interpolating among domains \cite{Bettoni:2018pbl}. The precise impact of the domain wall decay at symmetry restoration seems difficult to estimate without the use of lattice simulations. A naive extrapolation of the usual kink solution, strictly applicable only in the static regime, would suggest the gradient contribution to be commensurable with the potential energy density of the field at that time \cite{Saikawa:2017hiv}. Being this the case, the effective restoration of the symmetry due to backreaction effects will not generically translate into a smooth oscillatory pattern as that depicted in Fig.~\ref{fig1}, but rather into a non-perturbative stage where the created defects will percolate among themselves, releasing gradient energy, producing inhomogeneities and heating the system.  We postpone the study of this scenario to a future publication.

\section{Conclusions} \label{sec:conclusions}

The spontaneous breaking of discrete symmetries in the early decelerating Universe need not be a cosmological disaster. In this work, we have performed a detailed analysis of the early symmetry breaking dynamics of a non-minimally coupled $Z_2$-symmetric spectator field in quintessential inflation scenarios. The presence of a kinetic dominated era in this type of models triggers a Hubble-induced symmetry breaking for the spectator field soon after the end of inflation that forces its evolution towards large expectation values. The transition to the new minima proceeds through a tachyonic or spinodal instability able to extract energy from the gravitational field itself. The associated production of fluctuations in the infrared part of the spectrum turns out to be very efficient, giving rise to a semiclassical but stochastic picture where sizable and rather homogeneous field configurations emerge.
The homogeneity of these spatial domains is, however, a purely local concept since, as required by the model symmetries, the mean value of the scalar field remains zero at all times. From a global point of view, the Universe looks rather inhomogeneous, with regions of positive and negative field values separated by domain wall configurations.  These topological defects are, however, short-lived and present only till the onset of radiation domination or the appearance of sizable backreaction effects, whatever happens first. Beyond that time, the Hubble-induced symmetry breaking ceases effectively to exist and the domain walls are doomed to disappear.  The domain wall problem is consequently exorcised and replaced by a quite rich cosmological scenario. Among other interesting consequences, the temporal existence of topological defects may lead to the production of a detectable gravitational waves' background \cite{Bettoni:2018pbl} even in those cases in which the amplification of the primordial spectrum of gravitational waves during kinetic domination does not lead to any observable consequence ($\omega\simeq 1$) \cite{Opferkuch:2019zbd,Bernal:2019lpc,Figueroa:2019paj}. 
Our basic results can be applied to alternative symmetry breaking patterns leading to other topological defects such as strings or textures by simply replacing $\chi$ by its modulus. In particular, it would be interesting to apply our results to the Standard Model Higgs, extending the analysis of Ref.~\cite{Opferkuch:2019zbd} to non-homogeneous field configurations. 
Some work in refining the present treatment and the parameter space scanning is however necessary before extracting quantitative conclusions.  In particular, the results presented in this paper rely on simple analytical methods within the linear regime and should be consequently understood just as a preliminary step towards the proper characterization of the defects and their decay via lattice simulations. We believe, however, that the qualitative picture presented here is robust enough. 

\section*{Acknowledgments}

DB wishes to thank Jose Beltr\'an  for useful discussions. DB acknowledges support from the Attracc\'ion  del Talento  Cient\'ifico en Salamanca programme and from project PGC2018-096038-B-I00 by Spanish Ministerio de Ciencia, Innovac\'ion y Universidades. JR thanks Jose Beltr\'an and the University of Salamanca for the hospitality during the development of part of this project, as well as Mark Hindmarsh and Asier Lopez-Eiguren for useful discussions and Florencia A. Teppa Pannia for reference suggestions. 

\appendix
\bibliographystyle{JHEP.bst}
\bibliography{main6.bib}

\providecommand{\href}[2]{#2}\begingroup\raggedright\begin{thebibliography}{10}

\bibitem{Zeldovich:1974uw}
{\relax Ya}.~B. Zeldovich, I.~{\relax Yu}. Kobzarev and L.~B. Okun,
  \emph{{Cosmological Consequences of the Spontaneous Breakdown of Discrete
  Symmetry}}, {\emph{Zh. Eksp. Teor. Fiz.} {\bfseries 67} (1974) 3}.

\bibitem{Coulson:1995nv}
D.~Coulson, Z.~Lalak and B.~A. Ovrut, \emph{{Biased domain walls}},
  \href{https://doi.org/10.1103/PhysRevD.53.4237}{\emph{Phys. Rev.} {\bfseries
  D53} (1996) 4237}.

\bibitem{Preskill:1991kd}
J.~Preskill, S.~P. Trivedi, F.~Wilczek and M.~B. Wise, \emph{{Cosmology and
  broken discrete symmetry}},
  \href{https://doi.org/10.1016/0550-3213(91)90241-O}{\emph{Nucl. Phys.}
  {\bfseries B363} (1991) 207}.

\bibitem{Rai:1992xw}
B.~Rai and G.~Senjanovic, \emph{{Gravity and domain wall problem}},
  \href{https://doi.org/10.1103/PhysRevD.49.2729}{\emph{Phys. Rev.} {\bfseries
  D49} (1994) 2729} [\href{https://arxiv.org/abs/hep-ph/9301240}{{\ttfamily
  hep-ph/9301240}}].

\bibitem{McDonald:1997vy}
J.~McDonald, \emph{{Spontaneous discrete symmetry breaking during inflation and
  the NMSSM domain wall problem}},
  \href{https://doi.org/10.1016/S0550-3213(98)00414-3}{\emph{Nucl. Phys.}
  {\bfseries B530} (1998) 325}
  [\href{https://arxiv.org/abs/hep-ph/9709512}{{\ttfamily hep-ph/9709512}}].

\bibitem{Mazumdar:2015dwd}
A.~Mazumdar, K.~Saikawa, M.~Yamaguchi and J.~Yokoyama, \emph{{Possible
  resolution of the domain wall problem in the NMSSM}},
  \href{https://doi.org/10.1103/PhysRevD.93.025002}{\emph{Phys. Rev.}
  {\bfseries D93} (2016) 025002}
  [\href{https://arxiv.org/abs/1511.01905}{{\ttfamily 1511.01905}}].

\bibitem{Bettoni:2018pbl}
D.~Bettoni, G.~Dom\'enech and J.~Rubio, \emph{{Gravitational waves from global
  cosmic strings in quintessential inflation}},
  \href{https://doi.org/10.1088/1475-7516/2019/02/034}{\emph{JCAP} {\bfseries
  1902} (2019) 034} [\href{https://arxiv.org/abs/1810.11117}{{\ttfamily
  1810.11117}}].

\bibitem{Peebles:1998qn}
P.~J.~E. Peebles and A.~Vilenkin, \emph{{Quintessential inflation}},
  \href{https://doi.org/10.1103/PhysRevD.59.063505}{\emph{Phys. Rev.}
  {\bfseries D59} (1999) 063505}
  [\href{https://arxiv.org/abs/astro-ph/9810509}{{\ttfamily
  astro-ph/9810509}}].

\bibitem{Spokoiny:1993kt}
B.~Spokoiny, \emph{{Deflationary universe scenario}},
  \href{https://doi.org/10.1016/0370-2693(93)90155-B}{\emph{Phys. Lett.}
  {\bfseries B315} (1993) 40}
  [\href{https://arxiv.org/abs/gr-qc/9306008}{{\ttfamily gr-qc/9306008}}].

\bibitem{Wetterich:1987fm}
C.~Wetterich, \emph{{Cosmology and the Fate of Dilatation Symmetry}},
  \href{https://doi.org/10.1016/0550-3213(88)90193-9}{\emph{Nucl. Phys.}
  {\bfseries B302} (1988) 668}
  [\href{https://arxiv.org/abs/1711.03844}{{\ttfamily 1711.03844}}].

\bibitem{Wetterich:1994bg}
C.~Wetterich, \emph{{The Cosmon model for an asymptotically vanishing time
  dependent cosmological 'constant'}}, {\emph{Astron. Astrophys.} {\bfseries
  301} (1995) 321} [\href{https://arxiv.org/abs/hep-th/9408025}{{\ttfamily
  hep-th/9408025}}].

\bibitem{Wetterich:2014gaa}
C.~Wetterich, \emph{{Inflation, quintessence, and the origin of mass}},
  \href{https://doi.org/10.1016/j.nuclphysb.2015.05.019}{\emph{Nucl. Phys.}
  {\bfseries B897} (2015) 111}
  [\href{https://arxiv.org/abs/1408.0156}{{\ttfamily 1408.0156}}].

\bibitem{Hossain:2014xha}
M.~W. Hossain, R.~Myrzakulov, M.~Sami and E.~N. Saridakis, \emph{{Variable
  gravity: A suitable framework for quintessential inflation}},
  \href{https://doi.org/10.1103/PhysRevD.90.023512}{\emph{Phys. Rev.}
  {\bfseries D90} (2014) 023512}
  [\href{https://arxiv.org/abs/1402.6661}{{\ttfamily 1402.6661}}].

\bibitem{Rubio:2017gty}
J.~Rubio and C.~Wetterich, \emph{{Emergent scale symmetry: Connecting inflation
  and dark energy}},
  \href{https://doi.org/10.1103/PhysRevD.96.063509}{\emph{Phys. Rev.}
  {\bfseries D96} (2017) 063509}
  [\href{https://arxiv.org/abs/1705.00552}{{\ttfamily 1705.00552}}].

\bibitem{Dimopoulos:2017zvq}
K.~Dimopoulos and C.~Owen, \emph{{Quintessential Inflation with
  $\alpha$-attractors}},
  \href{https://doi.org/10.1088/1475-7516/2017/06/027}{\emph{JCAP} {\bfseries
  1706} (2017) 027} [\href{https://arxiv.org/abs/1703.00305}{{\ttfamily
  1703.00305}}].

\bibitem{Dimopoulos:2017tud}
K.~Dimopoulos, L.~Donaldson~Wood and C.~Owen, \emph{{Instant preheating in
  quintessential inflation with $\alpha$-attractors}},
  \href{https://doi.org/10.1103/PhysRevD.97.063525}{\emph{Phys. Rev.}
  {\bfseries D97} (2018) 063525}
  [\href{https://arxiv.org/abs/1712.01760}{{\ttfamily 1712.01760}}].

\bibitem{Akrami:2017cir}
Y.~Akrami, R.~Kallosh, A.~Linde and V.~Vardanyan, \emph{{Dark energy,
  $\alpha$-attractors, and large-scale structure surveys}},
  \href{https://doi.org/10.1088/1475-7516/2018/06/041}{\emph{JCAP} {\bfseries
  1806} (2018) 041} [\href{https://arxiv.org/abs/1712.09693}{{\ttfamily
  1712.09693}}].

\bibitem{Garcia-Garcia:2018hlc}
C.~García-García, E.~V. Linder, P.~Ruíz-Lapuente and M.~Zumalacárregui,
  \emph{{Dark energy from $\alpha$-attractors: phenomenology and observational
  constraints}},
  \href{https://doi.org/10.1088/1475-7516/2018/08/022}{\emph{JCAP} {\bfseries
  1808} (2018) 022} [\href{https://arxiv.org/abs/1803.00661}{{\ttfamily
  1803.00661}}].

\bibitem{Dimopoulos:2019gpz}
K.~Dimopoulos and L.~Donaldson-Wood, \emph{{Warm quintessential inflation}},
  \href{https://doi.org/10.1016/j.physletb.2019.07.017}{\emph{Phys. Lett.}
  {\bfseries B796} (2019) 26}
  [\href{https://arxiv.org/abs/1906.09648}{{\ttfamily 1906.09648}}].

\bibitem{Rosa:2019jci}
J.~G. Rosa and L.~B. Ventura, \emph{{Warm Little Inflaton becomes Dark
  Energy}}, \href{https://doi.org/10.1016/j.physletb.2019.134984}{\emph{Phys.
  Lett.} {\bfseries B798} (2019) 134984}
  [\href{https://arxiv.org/abs/1906.11835}{{\ttfamily 1906.11835}}].

\bibitem{Lima:2019yyv}
G.~B.~F. Lima and R.~O. Ramos, \emph{{Unified early and late Universe cosmology
  through dissipative effects in steep quintessential inflation potential
  models}},  \href{https://arxiv.org/abs/1910.05185}{{\ttfamily 1910.05185}}.

\bibitem{Ford:1986sy}
L.~H. Ford, \emph{{Gravitational Particle Creation and Inflation}},
  \href{https://doi.org/10.1103/PhysRevD.35.2955}{\emph{Phys. Rev.} {\bfseries
  D35} (1987) 2955}.

\bibitem{Felder:1999pv}
G.~N. Felder, L.~Kofman and A.~D. Linde, \emph{{Inflation and preheating in NO
  models}}, \href{https://doi.org/10.1103/PhysRevD.60.103505}{\emph{Phys. Rev.}
  {\bfseries D60} (1999) 103505}
  [\href{https://arxiv.org/abs/hep-ph/9903350}{{\ttfamily hep-ph/9903350}}].

\bibitem{Bettoni:2018utf}
D.~Bettoni and J.~Rubio, \emph{{Quintessential Affleck-Dine baryogenesis with
  non-minimal couplings}},
  \href{https://doi.org/10.1016/j.physletb.2018.07.046}{\emph{Phys. Lett.}
  {\bfseries B784} (2018) 122}
  [\href{https://arxiv.org/abs/1805.02669}{{\ttfamily 1805.02669}}].

\bibitem{Figueroa:2016dsc}
D.~G. Figueroa and C.~T. Byrnes, \emph{{The Standard Model Higgs as the origin
  of the hot Big Bang}},
  \href{https://doi.org/10.1016/j.physletb.2017.01.059}{\emph{Phys. Lett.}
  {\bfseries B767} (2017) 272}
  [\href{https://arxiv.org/abs/1604.03905}{{\ttfamily 1604.03905}}].

\bibitem{Nakama:2018gll}
T.~Nakama and J.~Yokoyama, \emph{{Reheating through the Higgs amplified by
  spinodal instabilities and gravitational creation of gravitons}},
  \href{https://doi.org/10.1093/ptep/ptz014}{\emph{PTEP} {\bfseries 2019}
  (2019) 033E02} [\href{https://arxiv.org/abs/1803.07111}{{\ttfamily
  1803.07111}}].

\bibitem{Dimopoulos:2018wfg}
K.~Dimopoulos and T.~Markkanen, \emph{{Non-minimal gravitational reheating
  during kination}},
  \href{https://doi.org/10.1088/1475-7516/2018/06/021}{\emph{JCAP} {\bfseries
  1806} (2018) 021} [\href{https://arxiv.org/abs/1803.07399}{{\ttfamily
  1803.07399}}].

\bibitem{Opferkuch:2019zbd}
T.~Opferkuch, P.~Schwaller and B.~A. Stefanek, \emph{{Ricci Reheating}},
  \href{https://doi.org/10.1088/1475-7516/2019/07/016}{\emph{JCAP} {\bfseries
  1907} (2019) 016} [\href{https://arxiv.org/abs/1905.06823}{{\ttfamily
  1905.06823}}].

\bibitem{Mukhanov:1990me}
V.~F. Mukhanov, H.~A. Feldman and R.~H. Brandenberger, \emph{{Theory of
  cosmological perturbations. Part 1. Classical perturbations. Part 2. Quantum
  theory of perturbations. Part 3. Extensions}},
  \href{https://doi.org/10.1016/0370-1573(92)90044-Z}{\emph{Phys. Rept.}
  {\bfseries 215} (1992) 203}.

\bibitem{Guth:1985ya}
A.~H. Guth and S.-Y. Pi, \emph{{The Quantum Mechanics of the Scalar Field in
  the New Inflationary Universe}},
  \href{https://doi.org/10.1103/PhysRevD.32.1899}{\emph{Phys. Rev.} {\bfseries
  D32} (1985) 1899}.

\bibitem{Felder:2000hj}
G.~N. Felder, J.~Garcia-Bellido, P.~B. Greene, L.~Kofman, A.~D. Linde and
  I.~Tkachev, \emph{{Dynamics of symmetry breaking and tachyonic preheating}},
  \href{https://doi.org/10.1103/PhysRevLett.87.011601}{\emph{Phys. Rev. Lett.}
  {\bfseries 87} (2001) 011601}
  [\href{https://arxiv.org/abs/hep-ph/0012142}{{\ttfamily hep-ph/0012142}}].

\bibitem{Felder:2001kt}
G.~N. Felder, L.~Kofman and A.~D. Linde, \emph{{Tachyonic instability and
  dynamics of spontaneous symmetry breaking}},
  \href{https://doi.org/10.1103/PhysRevD.64.123517}{\emph{Phys. Rev.}
  {\bfseries D64} (2001) 123517}
  [\href{https://arxiv.org/abs/hep-th/0106179}{{\ttfamily hep-th/0106179}}].

\bibitem{Copeland:2001qw}
E.~J. Copeland, D.~Lyth, A.~Rajantie and M.~Trodden, \emph{{Hybrid inflation
  and baryogenesis at the TeV scale}},
  \href{https://doi.org/10.1103/PhysRevD.64.043506}{\emph{Phys. Rev.}
  {\bfseries D64} (2001) 043506}
  [\href{https://arxiv.org/abs/hep-ph/0103231}{{\ttfamily hep-ph/0103231}}].

\bibitem{Asaka:2001ez}
T.~Asaka, W.~Buchmuller and L.~Covi, \emph{{False vacuum decay after
  inflation}}, \href{https://doi.org/10.1016/S0370-2693(01)00623-2}{\emph{Phys.
  Lett.} {\bfseries B510} (2001) 271}
  [\href{https://arxiv.org/abs/hep-ph/0104037}{{\ttfamily hep-ph/0104037}}].

\bibitem{GarciaBellido:2001cb}
J.~Garcia-Bellido and E.~Ruiz~Morales, \emph{{Particle production from symmetry
  breaking after inflation}},
  \href{https://doi.org/10.1016/S0370-2693(02)01820-8}{\emph{Phys. Lett.}
  {\bfseries B536} (2002) 193}
  [\href{https://arxiv.org/abs/hep-ph/0109230}{{\ttfamily hep-ph/0109230}}].

\bibitem{Copeland:2002ku}
E.~J. Copeland, S.~Pascoli and A.~Rajantie, \emph{{Dynamics of tachyonic
  preheating after hybrid inflation}},
  \href{https://doi.org/10.1103/PhysRevD.65.103517}{\emph{Phys. Rev.}
  {\bfseries D65} (2002) 103517}
  [\href{https://arxiv.org/abs/hep-ph/0202031}{{\ttfamily hep-ph/0202031}}].

\bibitem{GarciaBellido:2002aj}
J.~Garcia-Bellido, M.~Garcia~Perez and A.~Gonzalez-Arroyo, \emph{{Symmetry
  breaking and false vacuum decay after hybrid inflation}},
  \href{https://doi.org/10.1103/PhysRevD.67.103501}{\emph{Phys. Rev.}
  {\bfseries D67} (2003) 103501}
  [\href{https://arxiv.org/abs/hep-ph/0208228}{{\ttfamily hep-ph/0208228}}].

\bibitem{Polarski:1995jg}
D.~Polarski and A.~A. Starobinsky, \emph{{Semiclassicality and decoherence of
  cosmological perturbations}},
  \href{https://doi.org/10.1088/0264-9381/13/3/006}{\emph{Class. Quant. Grav.}
  {\bfseries 13} (1996) 377}
  [\href{https://arxiv.org/abs/gr-qc/9504030}{{\ttfamily gr-qc/9504030}}].

\bibitem{Lesgourgues:1996jc}
J.~Lesgourgues, D.~Polarski and A.~A. Starobinsky, \emph{{Quantum to classical
  transition of cosmological perturbations for nonvacuum initial states}},
  \href{https://doi.org/10.1016/S0550-3213(97)00224-1}{\emph{Nucl. Phys.}
  {\bfseries B497} (1997) 479}
  [\href{https://arxiv.org/abs/gr-qc/9611019}{{\ttfamily gr-qc/9611019}}].

\bibitem{Kiefer:1998jk}
C.~Kiefer and D.~Polarski, \emph{{Emergence of classicality for primordial
  fluctuations: Concepts and analogies}},
  \href{https://doi.org/10.1002/andp.2090070302}{\emph{Annalen Phys.}
  {\bfseries 7} (1998) 137}
  [\href{https://arxiv.org/abs/gr-qc/9805014}{{\ttfamily gr-qc/9805014}}].

\bibitem{abramowitz+stegun}
M.~Abramowitz and I.~A. Stegun, \emph{Handbook of Mathematical Functions with
  Formulas, Graphs, and Mathematical Tables}. Dover, New York, ninth dover
  printing, tenth gpo printing~ed., 1964.

\bibitem{Boyanovsky:1993xf}
D.~Boyanovsky, H.~J. de~Vega and R.~Holman, \emph{{Nonequilibrium evolution of
  scalar fields in FRW cosmologies I}},
  \href{https://doi.org/10.1103/PhysRevD.49.2769}{\emph{Phys. Rev.} {\bfseries
  D49} (1994) 2769} [\href{https://arxiv.org/abs/hep-ph/9310319}{{\ttfamily
  hep-ph/9310319}}].

\bibitem{Cooper:1994hr}
F.~Cooper, S.~Habib, Y.~Kluger, E.~Mottola, J.~P. Paz and P.~R. Anderson,
  \emph{{Nonequilibrium quantum fields in the large N expansion}},
  \href{https://doi.org/10.1103/PhysRevD.50.2848}{\emph{Phys. Rev.} {\bfseries
  D50} (1994) 2848} [\href{https://arxiv.org/abs/hep-ph/9405352}{{\ttfamily
  hep-ph/9405352}}].

\bibitem{Cooper:1996ii}
F.~Cooper, S.~Habib, Y.~Kluger and E.~Mottola, \emph{{Nonequilibrium dynamics
  of symmetry breaking in lambda Phi**4 field theory}},
  \href{https://doi.org/10.1103/PhysRevD.55.6471}{\emph{Phys. Rev.} {\bfseries
  D55} (1997) 6471} [\href{https://arxiv.org/abs/hep-ph/9610345}{{\ttfamily
  hep-ph/9610345}}].

\bibitem{Baacke:1996se}
J.~Baacke, K.~Heitmann and C.~Patzold, \emph{{Nonequilibrium dynamics: A
  Renormalized computation scheme}},
  \href{https://doi.org/10.1103/PhysRevD.55.2320}{\emph{Phys. Rev.} {\bfseries
  D55} (1997) 2320} [\href{https://arxiv.org/abs/hep-th/9608006}{{\ttfamily
  hep-th/9608006}}].

\bibitem{Baacke:1997rs}
J.~Baacke, K.~Heitmann and C.~Patzold, \emph{{Renormalization of nonequilibrium
  dynamics in FRW cosmology}},
  \href{https://doi.org/10.1103/PhysRevD.56.6556}{\emph{Phys. Rev.} {\bfseries
  D56} (1997) 6556} [\href{https://arxiv.org/abs/hep-ph/9706274}{{\ttfamily
  hep-ph/9706274}}].

\bibitem{Boyanovsky:1998yp}
D.~Boyanovsky, H.~J. de~Vega, R.~Holman and J.~Salgado, \emph{{Nonequilibrium
  Bose-Einstein condensates, dynamical scaling and symmetric evolution in the
  large N Phi**4 theory}},
  \href{https://doi.org/10.1103/PhysRevD.59.125009}{\emph{Phys. Rev.}
  {\bfseries D59} (1999) 125009}
  [\href{https://arxiv.org/abs/hep-ph/9811273}{{\ttfamily hep-ph/9811273}}].

\bibitem{Bardeen:1985tr}
J.~M. Bardeen, J.~R. Bond, N.~Kaiser and A.~S. Szalay, \emph{{The Statistics of
  Peaks of Gaussian Random Fields}},
  \href{https://doi.org/10.1086/164143}{\emph{Astrophys. J.} {\bfseries 304}
  (1986) 15}.

\bibitem{Halperin}
B.~Halperin, \emph{{Physics of Defects}}. R.~Balian, M.~Kleman, and
  J.~P.~Poirier, North-Holland, New York, 1981.

\bibitem{Liu:1992zz}
F.~Liu and G.~F. Mazenko, \emph{{Defect-defect correlation in the dynamics of
  first-order phase transitions}},
  \href{https://doi.org/10.1103/PhysRevB.46.5963}{\emph{Phys. Rev.} {\bfseries
  B46} (1992) 5963}.

\bibitem{Saikawa:2017hiv}
K.~Saikawa, \emph{{A review of gravitational waves from cosmic domain walls}},
  \href{https://doi.org/10.3390/universe3020040}{\emph{Universe} {\bfseries 3}
  (2017) 40} [\href{https://arxiv.org/abs/1703.02576}{{\ttfamily 1703.02576}}].

\bibitem{Bernal:2019lpc}
N.~Bernal and F.~Hajkarim, \emph{{Primordial Gravitational Waves in Nonstandard
  Cosmologies}}, \href{https://doi.org/10.1103/PhysRevD.100.063502}{\emph{Phys.
  Rev.} {\bfseries D100} (2019) 063502}
  [\href{https://arxiv.org/abs/1905.10410}{{\ttfamily 1905.10410}}].

\bibitem{Figueroa:2019paj}
D.~G. Figueroa and E.~H. Tanin, \emph{{Ability of LIGO and LISA to probe the
  equation of state of the early Universe}},
  \href{https://doi.org/10.1088/1475-7516/2019/08/011}{\emph{JCAP} {\bfseries
  1908} (2019) 011} [\href{https://arxiv.org/abs/1905.11960}{{\ttfamily
  1905.11960}}].

\end{thebibliography}\endgroup
\end{document}